\documentclass[a4paper,american,aps,citeautoscript,floatfix,longbibliography,pdftex,pre,superscriptaddress,showpacs,preprint]{revtex4-1}
\usepackage{amssymb, graphicx,subfigure}
\usepackage{enumerate,tensor}
\usepackage{amsmath,bm}
\usepackage{dcolumn}
\usepackage{color}
\usepackage{natbib} 
\usepackage[latin1]{inputenc}
\usepackage{subfigure}
\usepackage{textcomp}
\usepackage{hyperref}
\usepackage{booktabs}
\usepackage{makecell}
\usepackage{psfrag}
\usepackage{epsfig}

\newcommand{\eg}{e.\,g.}%
\newcommand{\ie}{i.\,e.}%

\makeatletter
\def\paragraph{\@startsection{paragraph}{4}{10pt}{-1.25ex plus -1ex minus -.1ex}{0ex plus 0ex}{\normalsize\textit}}
\renewcommand\@biblabel[1]{#1}
\renewcommand\@makefntext[1]%
{\noindent\makebox[0pt][r]{\@thefnmark\,}#1}
\DeclareRobustCommand\onlinecite{\@onlinecite}
\def\@onlinecite#1{\begingroup\let\@cite\NAT@citenum\citealp{#1}\endgroup}
\def\tagform@#1{\maketag@@@{\ignorespaces#1\unskip\@@italiccorr}}
\let\orgtheequation\theequation
\def\theequation{(\orgtheequation)}
\makeatother

\begin{document}

\title{Nonlinear dynamics of atoms in a crossed optical dipole trap}

\author{Rosario Gonz\'alez-F\'erez}
%\email{rogonzal@ugr.es}%
\affiliation{Instituto Carlos I de F\'{\i}sica Te\'orica y Computacional,
and Departamento de F\'{\i}sica At\'omica, Molecular y Nuclear,
  Universidad de Granada, 18071 Granada, Spain} 
\affiliation{The Hamburg Center for Ultrafast Imaging, Luruper Chaussee 149, 22761 Hamburg, Germany}

\author{Manuel I\~narrea}
\affiliation{\'Area de F\'{\i}sica, Universidad de la Rioja, 26006 Logro\~no, La Rioja, Spain}

  \author{J. Pablo Salas}
\affiliation{\'Area de F\'{\i}sica, Universidad de la Rioja, 26006 Logro\~no, La Rioja, Spain}

\author{Peter Schmelcher}
\affiliation{The Hamburg Center for Ultrafast Imaging, Luruper Chaussee 149, 22761 Hamburg, Germany}
\affiliation{Zentrum f\"ur Optische Quantentechnologien, Universit\"at
  Hamburg, Luruper Chaussee 149, 22761 Hamburg, Germany}

\date{\today}
\begin{abstract} 
We explore the classical dynamics of  atoms in an optical dipole trap formed by two identical Gaussian beams propagating 
in perpendicular directions. The phase space is a mixture of regular and  chaotic orbits, the later becoming dominant as the energy of the
 atoms increases.
The trapping capabilities of these perpendicular Gaussian beams  are investigated by considering 
an atomic ensemble in free motion. 
After a sudden turn on of the dipole trap, a certain fraction of atoms  in the ensemble  remains  trapped.  The majority of these trapped  atoms has energies larger than the  escape channels, which can be 
 explained by the existence of 
 regular  and chaotic orbits with very long escape times. 
\end{abstract}
\maketitle

\section{Introduction}
\label{sec:introduction}
Since the seventies, it is well known that the  interaction
between the induced atomic dipole moment and the
intensity gradient of a non-resonant light
enables the optical confinement  of neutral atoms \cite{ashkin,varios,cable,grimm,MetcalfStraten}.
Nowadays, these optical dipole traps are  routinely used to confine neutral atoms in the
cold and ultracold regime where quantum-engineering is required.  These confined  atomic ensembles
allow for a wide range of applications  such as 
single atom manipulation~\cite{PhysRevA.67.033403}, 
Bose-Einstein condensation \cite{bec}, optical atomic clocks \cite{clocks} or the observation of classical and quantum
chaos~\cite{PhysRevLett.86.1514,PhysRevLett.86.1518}.

%\medskip\noindent
The simplest optical trap providing confinement of neutral atoms consists of a single
strongly focused Gaussian laser beam \cite{ashkin,cable}. 
In this case, the confinement is one-dimensional and perpendicular to the propagation of the beam axis.
For a tight confinement in the three spatial dimensions,  the so-called crossed-beam trap is
commonly used which consists of two
Gaussian laser beams with orthogonal polarizations  propagating along perpendicular directions
 \cite{MetcalfStraten,grimm,PhysRevLett.74.3577}.
With this trap, it
is possible to obtain highly isotropic atomic ensembles
tightly confined in all dimensions~\cite{MetcalfStraten,grimm,PhysRevLett.74.3577}.

Opposite to a Bose-Einstein condensate, for a thermal atomic cloud, quantum effects can be neglected and it is appropriate to
describe  the optical trapping
mechanism via the corresponding classical dynamics of the atoms in the electromagnetic fields. 
 Indeed, the non-linear nature of the optical trapping renders these systems very attractive for   
classical studies. 
At the same time, such classical dynamics studies  are scarce in the literature. 
In this sense,  Barker and coworkers \cite{barker} used a 
classical one-dimensional model to explain the Stark deceleration of a cold molecular beam by a single focussed laser beam. 
The main goal of this paper is to perform a systematic study of the classical dynamics of optically trapped neutral atoms.
We also investigate the phase space of an atomic ensemble in free motion 
that is exposed to a suddenly switched on
crossed optical dipole trap, and show that a certain fraction of the atoms are indeed trapped. 
The standard procedure of trapping neutral atoms relies on adiabatic processes, 
 meaning that one first slows or even stops a particle beam and consequently  traps it. 
 The sudden trapping mechanism investigated here represents a highly non-adiabatic and instantaneous process.
However,  due to its in particular experimental simplicity (there is no need of cooling techniques), this trapping procedure could be of interest.

\medskip\noindent
The paper is organized as follows: In \autoref{sec:hamiltonian} we establish the three degree of
freedom Hamiltonian governing the dynamics of an atom in a crossed-beam trap.
The study of the critical points of this system  and the description of the fundamental families of periodic orbits 
are also provided in this section.
We explore   the nonlinear dynamics of  the system by means of a fast chaos indicator in \autoref{dynamics}.  
In \autoref{sec:trap2d}, we
investigate the  dynamics of an atomic beam in free motion which is suddenly exposed to a crossed-beam trap. 
The conclusions are provided in \autoref{sec:conclusions}.

\section{Classical Hamiltonian of a single atom in an optical dipole trap}
\label{sec:hamiltonian}

\begin{figure}
\centerline{\includegraphics[scale=0.38]{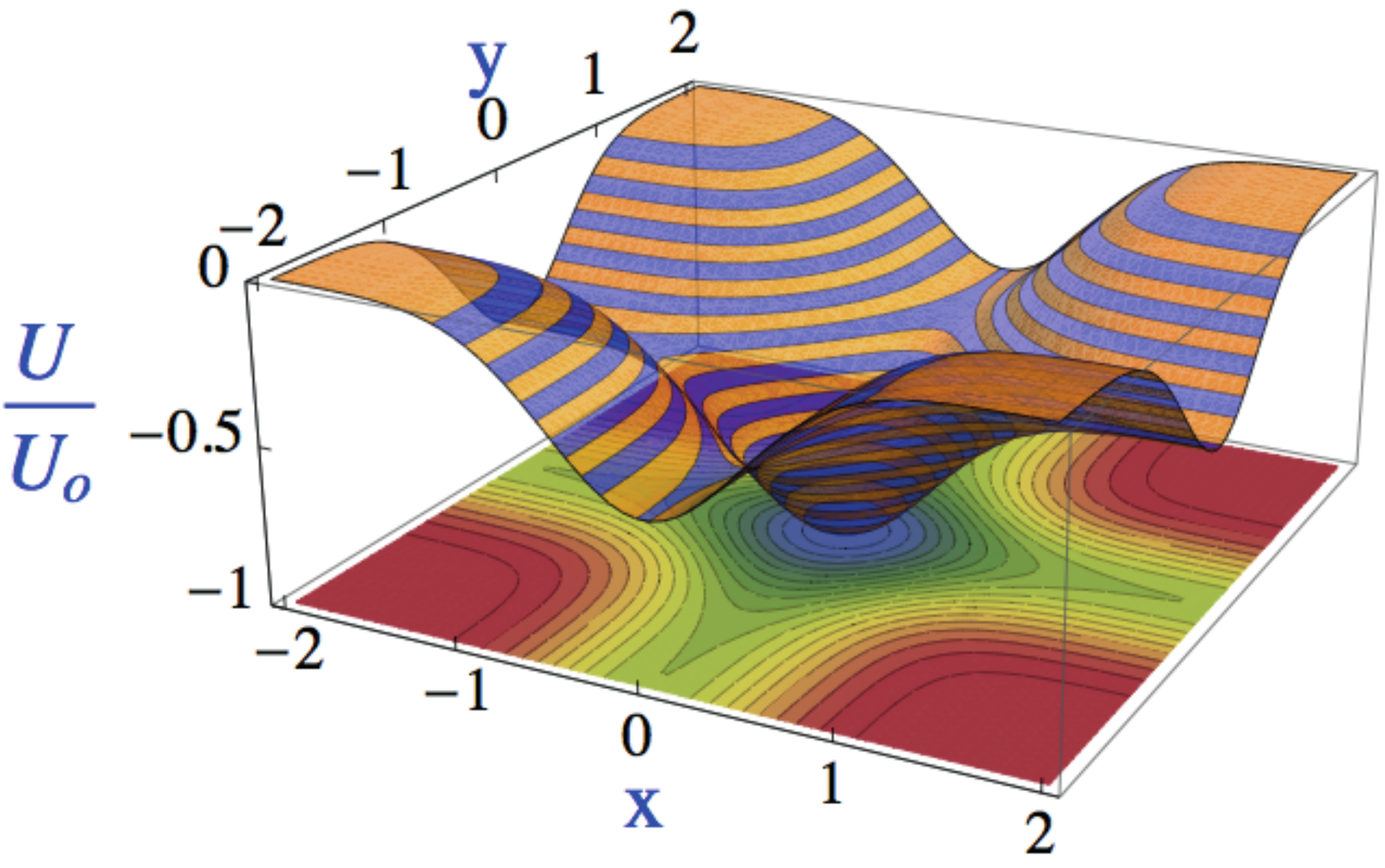}} 

\centerline{\includegraphics[scale=0.38]{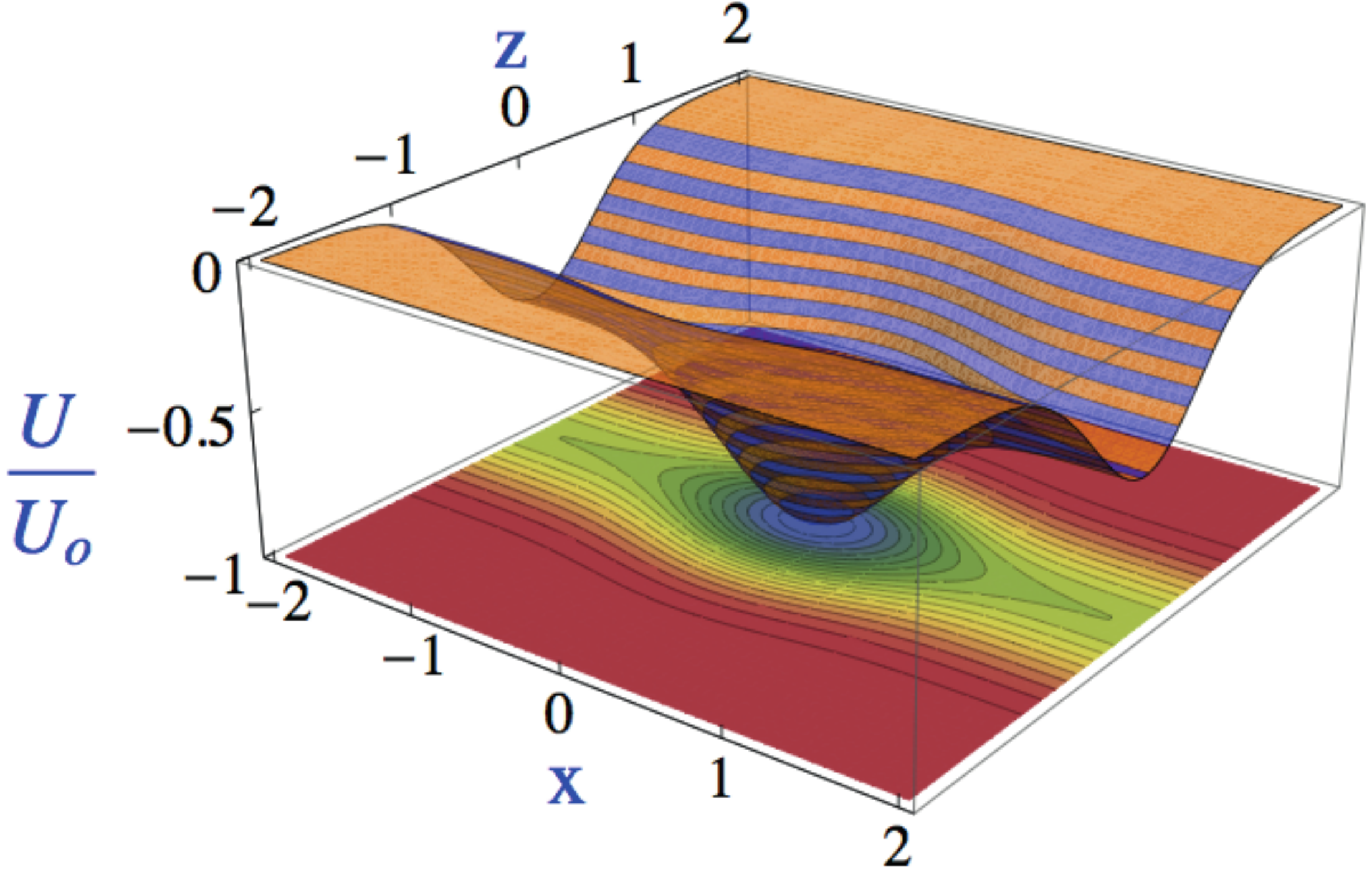}} 
\caption{Upper panel: Interaction potential $u(x,y,z)$ along the $z=0$ plane. Lower panel: 
Interaction potential $u(x,y,z)$ along the $y=0$ plane. Dimensionless
coordinates $(x=X/\omega_o,y=Y/\omega_o,z=Z/\omega_o)$ are used.}
\label{fi:poten1}
\end{figure}

When an atom is exposed to laser light, the electric field $\vec E$ of the laser
induces a dipole moment $\vec d$  in the atom given by $\vec d=\alpha(\omega) \ \vec E $, 
where $ \alpha(\omega)$ is the atomic polarizability which depends on the laser driving frequency  $\omega$~\cite{grimm}. 
%that oscillates with the laser 
%frequency $\Omega$. 
%%The induced dipole moment is given by
%%\[
%%\vec d = \alpha(\Omega) \ \vec E,
%%\]
%%\noindent
%which depends on $\Omega$. 
For the non-resonant case, the frequency of the non-resonant light
is assumed to be far detuned from any atomic transitions.
Using the rotating wave approximation~\cite{PershanPR66}, 
%In a non resonant situation \cite{grimm,PershanPR66}, the average
%over the fast laser frequency $\Omega$
%allows one  
the interaction
potential $U$ of the dipole $\vec d$ in the field $\vec E$ reads as~\cite{grimm}
\begin{equation}
\label{dipole}
U=-\frac{1}{2} \ \langle\vec d \cdot \vec E\rangle,
\end{equation}
where the brackets denote the time average of the terms depending on the frequency $\omega$~\cite{PershanPR66}.
Here, we consider an atom exposed to two identical focused Gaussian laser beams, which  propagate 
along the perpendicular axes $X$ and $Y$, and that are polarized along the (perpendicular)
directions $Z$ and $X$,  respectively.
The total electric field of the two beams is
\begin{eqnarray}
\label{field}
\nonumber
\vec E(X,Y,Z)=& &E_o  \ \exp\left[-\frac{Y^2+Z^2}{\omega_o^2}\right] \cos(\omega t)\widehat X\\
& + & E_o  \ \exp\left[-\frac{X^2+Z^2}{\omega_o^2}\right] \cos(\omega t)\widehat Y
\end{eqnarray}
where $E_o$ and $\omega_o$ are the electric field strength and the waist of 
the beams, respectively.  Thence, the dipole potential~\ref{dipole}
becomes
\begin{eqnarray}
\label{poten}
U(X,Y,Z)=&-&\frac{U_o}{2} \exp\left[-\frac{2 \, (Y^2+Z^2)}{\omega_o^2}\right]\\
&-&
\frac{U_o}{2} \exp\left[-\frac{2 \, (X^2+Z^2)}{\omega_o^2}\right], 
\nonumber
\end{eqnarray}
\noindent
where $U_o=\alpha_oE_o^2/2$, with 
$\alpha_o$ being the average atomic polarizability.
In  \autoref{field}-\ref{poten},  we are assuming that the beam waist $\omega_o$ of the laser field is 
much larger than its wavelength   $\omega_o \gg \lambda$~\cite{grimm}.
The dipole potential \ref{poten} has a critical point at the origin with a minimum energy $-U_o$, 
its effective depth is $U_o/2$ along the $X$ and $Y$ axes,  and $U_o$ along the 
$Z$ direction. This is
illustrated in Fig. \ref{fi:poten1}, where  the characteristic
exchange symmetry between the coordinates $x$ and $y$ is observed.

\medskip
The classical Hamiltonian describing the motion of an atom of
mass $m$ in this cross-beam trap is given by
\begin{equation}
\label{ham}
{\cal H} = \frac{P_X^2+P_Y^2+P_Z^2}{2 m} +U(X,Y,Z),
\end{equation}
\noindent
where $V_i=P_i/m$, $i=X, Y, Z$,  are the Cartesian components of
the velocity of the atom. For this study, it is useful to introduce a dimensionless version of the Hamiltonian
\ref{ham}. To do so, we define the dimensionless coordinates 
$x=X/w_o$, $y = Y/w_o$ and $z=Z/w_o$ and  the dimensionless unit of time $t'=t \ \nu$
with the frequency $\nu=(U_o/w_o^2 m)^{1/2}$.  Applying
these transformations to the Hamiltonian \ref{ham},  we get the following dimensionless Hamiltonian
\begin{equation}
\label{ham1}
E={\cal H} = \frac{v_x^2+v_y^2+v_z^2}{2}+u(x,y,z),
\end{equation}
\noindent
where $v_i=dx_i/dt$ are the corresponding velocity components of the atom
and the (dimensionless) dipole potential $u(x,y,z)$ takes the form
\begin{eqnarray}
\label{poten2}
u(x,y,z)=&-&\frac{1}{2} 
%\ \left\{\right. 
\exp\left[-2 \, (y^2+z^2)\right]\\
&-&
\frac{1}{2} \exp\left[-2 \, (x^2+z^2)\right].%\left.\right\}, 
\nonumber
\end{eqnarray}
\noindent
The coordinates and the energy $E$ are given in units of $w_o$ and 
$U_o$, respectively. Thence, the energy of the  dipole trap at its minimum is  $-1$ and its effective depth is $-1/2$
(see \autoref{fi:poten1}).

\medskip\par
\begin{figure}[h]
\centerline{\includegraphics[scale=0.5]{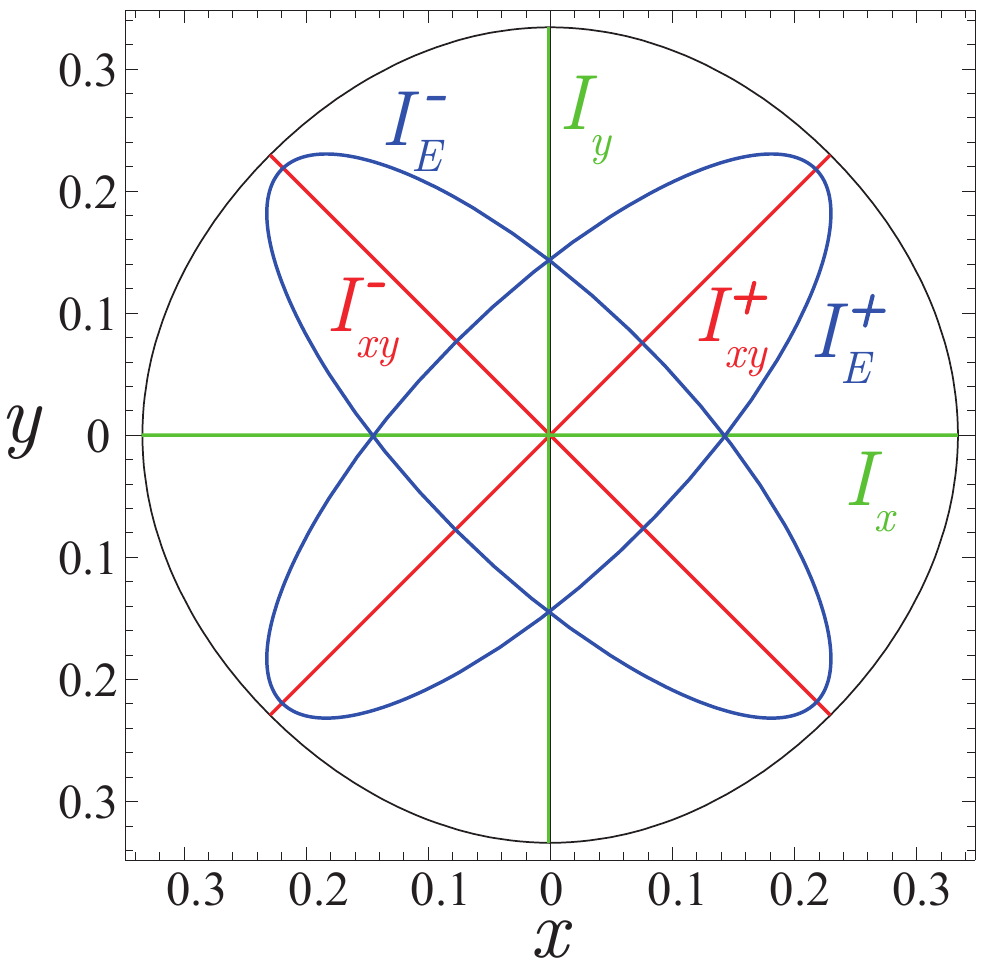}} 
\caption{Examples  of the periodic orbits $I_x$, $I_y$, $I_{xy}^+$, $I_{xy}^-$, $I_{E}^+$ and 
$I_E^-$ in the $xy$-plane for the energy $E=-0.9$.}
\label{fi:analiticas}
\end{figure}

The classical  equations of motion read 
\begin{eqnarray}
\label{ecumovi}
\dot x &=& v_x, \qquad \dot y = v_y, \qquad \dot z = v_z, \nonumber \\[2ex] 
\dot v_x &=& -2 \ x \ \exp\left[-2 (x^2+z^2)\right], \label{movi3D}\\[2ex]
\dot v_y &=& -2 \ y \ \exp\left[-2 (y^2+z^2)\right], \nonumber \\[2ex]
\dot v_z &=&-2 \ z \ \left\{ \exp\left[-2 (y^2+z^2)\right]+
\exp\left[-2 (x^2+z^2)\right]\right\} \nonumber.
\end{eqnarray}
In this system of coupled equations, there are  three families of periodic orbits. 
The rectilinear orbits, $I_x$, $I_y$ and $I_z$,  along the $x$, $y$ and $z$ axes,
which are always particular solutions of~\autoref{ecumovi}.
Orbits $I_x$ and $I_y$ are plotted in~\autoref{fi:analiticas}. 
For $z=v_z=0$, we find the rectilinear orbits,  $I_{xy}^\pm$, with initial conditions 
$x = \pm y$ and $v_x=\pm v_y$, \ie, the two bisectors of the $xy$ plane,
and the elliptic-shaped trajectories, $I_E^\pm$, 
 for the initial conditions $x = \pm y$ and $v_x=\mp v_y$ (see \autoref{fi:analiticas}). 

%\noindent
In the invariant plane $z=v_z=0$, \ie, the trajectories with initial conditions $z(0)=v_z(0)=0$ always satisfy 	
$z(t)=v_z(t)=0$. The degrees of freedom $(x,v_x)$ and $(y,v_y)$ are 
decoupled in the Hamiltonian~\eqref{ham1} and the system is integrable.
There are two more invariant planes given by $x=v_x=0$ and  $y=v_y=0$
which present an equivalent dynamics with two degrees of freedom due to the symmetry between the 
coordinates $x$ and $y$, and  between their corresponding velocities.
If the motion of the atom is confined close to the trap minimum, the dipole potential \eqref{poten2} could be 
approximated by the harmonic one $U(x,y,z) \approx x^2+y^2+2 z^2-1$, and the system becomes a
harmonic oscillator in 3D which is integrable and separable.

\section{Nonlinear Dynamics of  the trapped atom}
\label{dynamics}
Here, we focus on the phase space structure of an atom confined in the optical dipole trap
governed by the Hamiltonian \ref{ham1}. The latter involves three degrees of freedom, and
the corresponding  phase space is six-dimensional leading to a five-dimensional energy shell.
Thus, Poincar\'e  surfaces of section \cite{poincare,poincare1} are not
useful to investigate  the phase space structures.  
The  so-called chaos indicators represent an alternative  to perform such a study~\cite{barrio,barrio1}.
One of the most  popular is the Orthogonal Fast
Lyapunov indicator (OFLI)~\cite{ofli1,ofli2,ofli22}, 
defined as
\begin{equation}
\label{def:ofli}
\mbox{OFLI}({\bf r}_0, \delta{\bf r}_0, t_f) = \sup_{0 \le t \le t_f} \log ||\delta{\bf r}^\bot(t)||
\end{equation}
where  ${\bf r}(t) =(x(t),y(t),z(t))$, ${\bf r}_o={\bf r}(0) $, 
$\delta{\bf r}^\bot$ is the component of the variational vector
 $\delta{\bf y}$ orthogonal to the flow $d {\bf r}/dt$,  and $t_f$ is the stopping time. 
The OFLI provides a fast way to determine if an orbit is chaotic
and is able to distinguish between periodic and resonant orbits. In this way, the variational vector
$\delta{\bf r}^\bot$ behaves
linearly for regular resonant orbits and for orbits on a KAM torus;  it tends to constant
values for periodic ones,  and it increases exponentially for chaotic ones \cite{ofli2,ofli22,ofli3}.
 Examples of this behavior are shown in Fig. 4 and Fig. 5.2 of  references
 \cite{barrio} and \cite{ofli33}, respectively.
Note that the Lyapunov exponents are  defined in the long-time limit (see e.g.  Ref.~\cite{friedrich}), 
in contrast to the  OFLI definition \eqref{def:ofli}. 
The main disadvantage   is that the OFLI results depend on the initial conditions of the
variational vector $\delta{\bf y}_o$.
Here, we use 
the so-called OFLI$^{TT}_2$ extension of OFLI~\cite{barrio,barrio1},  
which  removes the
drawback of choosing $\delta{\bf y}_o$ by incorporating the
second order variational equations in the computation of the indicator.

In the  six-dimensional system \eqref{ham1}, a two-dimensional OFLI$^{TT}_2$ map is obtained by 
imposing three restrictions in addition of fixing the energy $E$.
To implement the latter, we have to choose two proper two-dimensional subspaces of initial conditions. 
Since  the $x$ and $y$ coordinates are dynamically equivalent,  
we choose the two planar subspaces
 ${\cal S}_1 =\left\{(y,z),x=v_y=v_z=0\right\}$ and ${\cal S}_2 =\left\{(x,y),z=v_x=v_y=0\right\}$, 
 which are perpendicular to the $x$ and $z$-axes, respectively.
For a fixed energy  $E$, the energy condition \eqref{ham1} provides the initial value for  the velocities
\begin{eqnarray}
\label{regions1}
v_x &=& \pm \sqrt{2 E + \exp[-2 (y^2+z^2)]+\exp[-2 z^2]},\\[2ex]
\label{regions2}
v_z &=& \pm \sqrt{2 E + \exp[-2 x^2]+\exp[-2 y^2]},
\end{eqnarray}
 in ${\cal S}_1$ and ${\cal S}_2$,  respectively.
For $v_x=0$ and $v_z=0$, these equations give the  available
regions $(y,z)$ and $(x,y)$ in the planar subspaces ${\cal S}_1$ and ${\cal S}_2$,  respectively.

%The periodic orbits $I_x$, $I_y$ and $I_z$ correspond to the central point $(0,0)$ in  ${\cal S}_1$
%and ${\cal S}_2$, respectively;
%$I_x$ ($I_y$) appears at the rightmost (leftmost) points in the $y=0$ ($z=0$)-axis of ${\cal S}_2$ 
%(${\cal S}_1$); and  $I_y$ ($I_z$) is at the upper (lower) end points of the $x=0$ ($y=0$)-axis 
%in ${\cal S}_2$ (${\cal S}_1$). 

\begin{figure}[t]
\centerline{\includegraphics[scale=0.4]{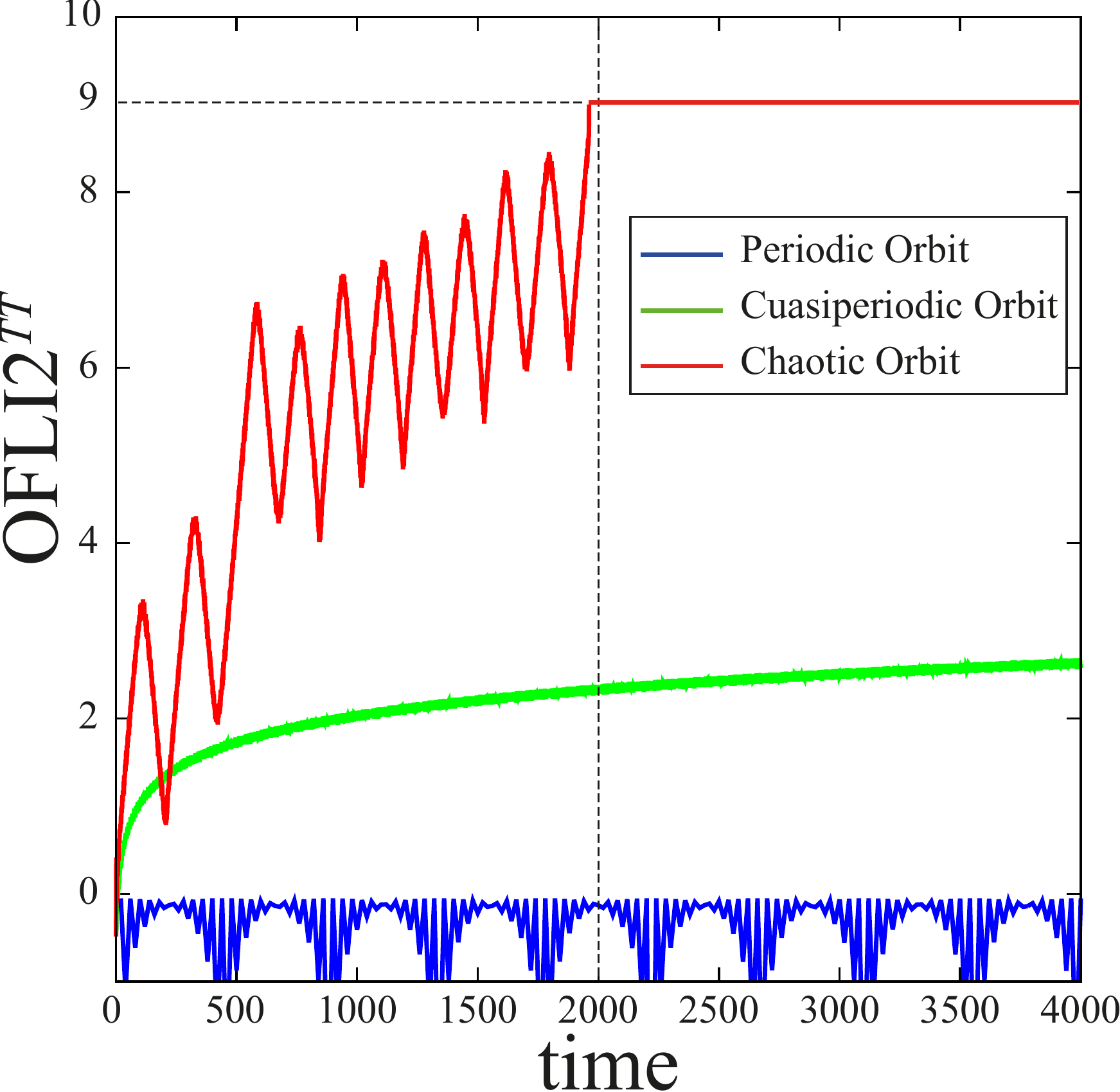} \quad \includegraphics[scale=0.4]{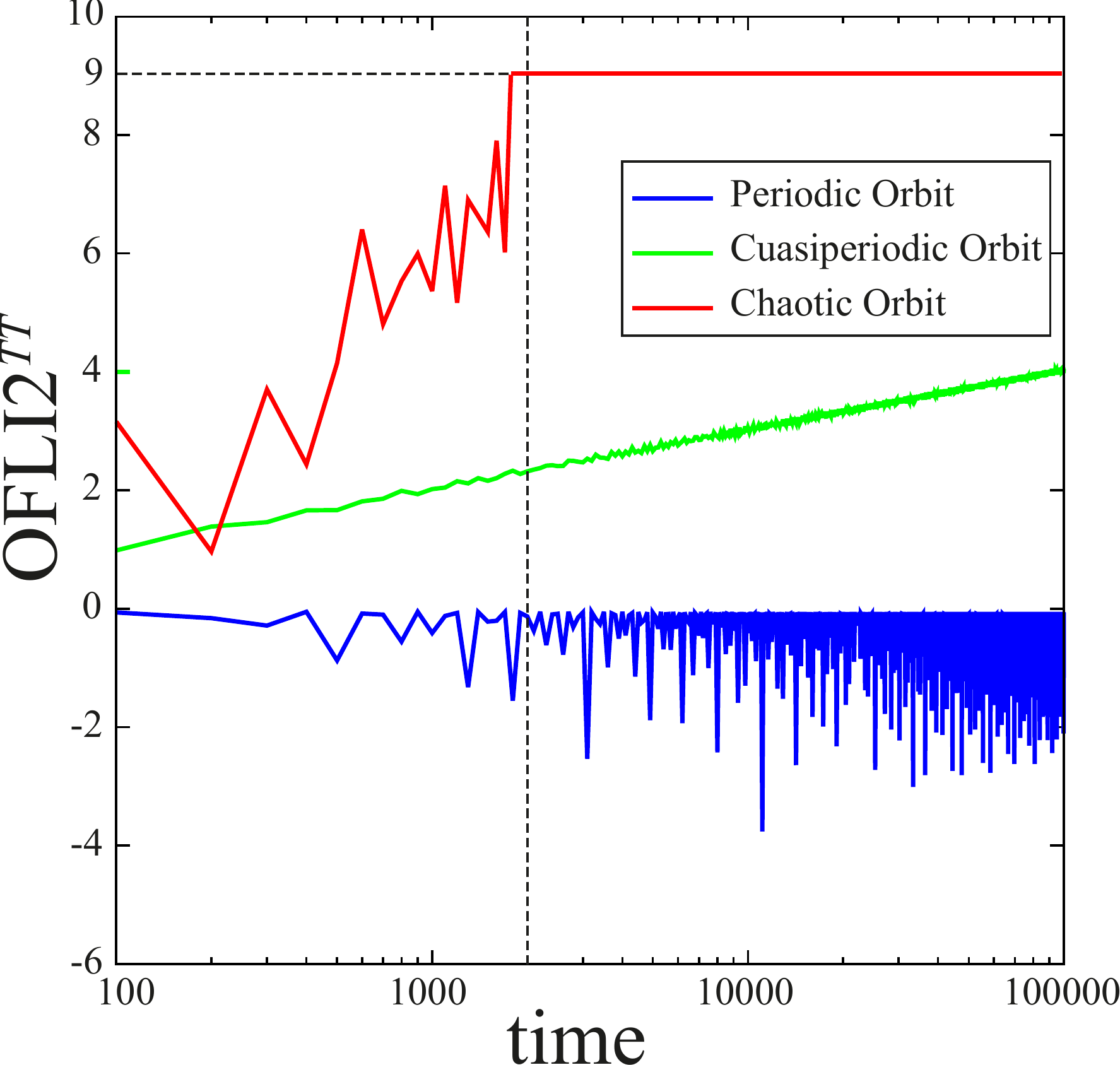}}
\caption{Sort (left panel) and long (right panel) time
evolution of the OFLI$^{TT}_2$ indicator for a quasiperiodic and  a chaotic
trajectories and for the periodic orbit $I_z$. 
Note that in the right panel a logarithmic scale has been used for the abscissa axis. 
Their energy is $E=-0.6$  and their initial conditions
have  been taken from the OFLI$^{TT}_2$ map of Fig.\ref{fi:ofli2_06}~b.
The vertical and horizontal dashed lines
indicate the stopping time $t_f=2000$ and the cut-off value nine of the OFLI$^{TT}_2$ for 
chaotic orbits, respectively.}
\label{fi:ofli2_time}
\end{figure}

We have calculated the OFLI$^{TT}_2$ by the numerical integration of the Hamiltonian equations of motion \ref{ecumovi} using an explicit 
Runge-Kutta algorithm of eighth order with step size control and dense output~\cite{Hairer}.
In our calculations, we stop the time evolution, if the OFLI$^{TT}_2$ reaches the value nine that characterizes
a chaotic orbit or if $t$ becomes larger than the stopping time $t_f =2000$.
Our numerical tests have shown that this value for the stopping 
time is appropriate for the correct characterization of any orbit.
In Fig.~\ref{fi:ofli2_time} we present 
 the short and the long time evolution of the OFLI$^{TT}_2$ for a quasiperiodic and a
chaotic trajectories and for the periodic orbit $I_z$. The energy of these orbits is $E=-0.6$ and their initial conditions
have been taken from the OFLI$^{TT}_2$ map of Fig.~\ref{fi:ofli2_06}~b.
 These time evolutions  show that with  a stopping time of $t_f =2000$ the main features of these orbits are correctly captured.

In the OFLI$^{TT}_2$ maps, 
we have used a RGB (Red Green Blue) color code: blue and red  are  associated to regular and chaotic  orbits, respectively; 
intermediate colors from dark blue to red indicate the evolution from regular to chaotic motion; finally pink and white
 stand for not allowed initial conditions and  escape orbits,  respectively.

 \begin{figure}[t]
\centerline{\includegraphics[scale=0.7]{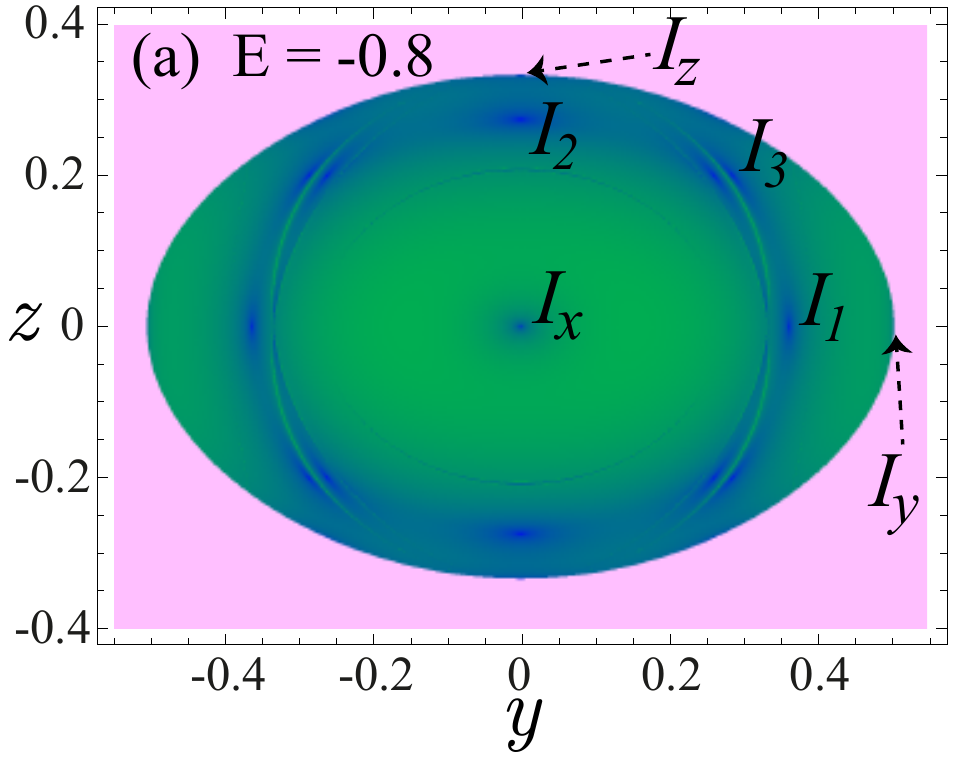} }
 \centerline{\includegraphics[scale=0.7]{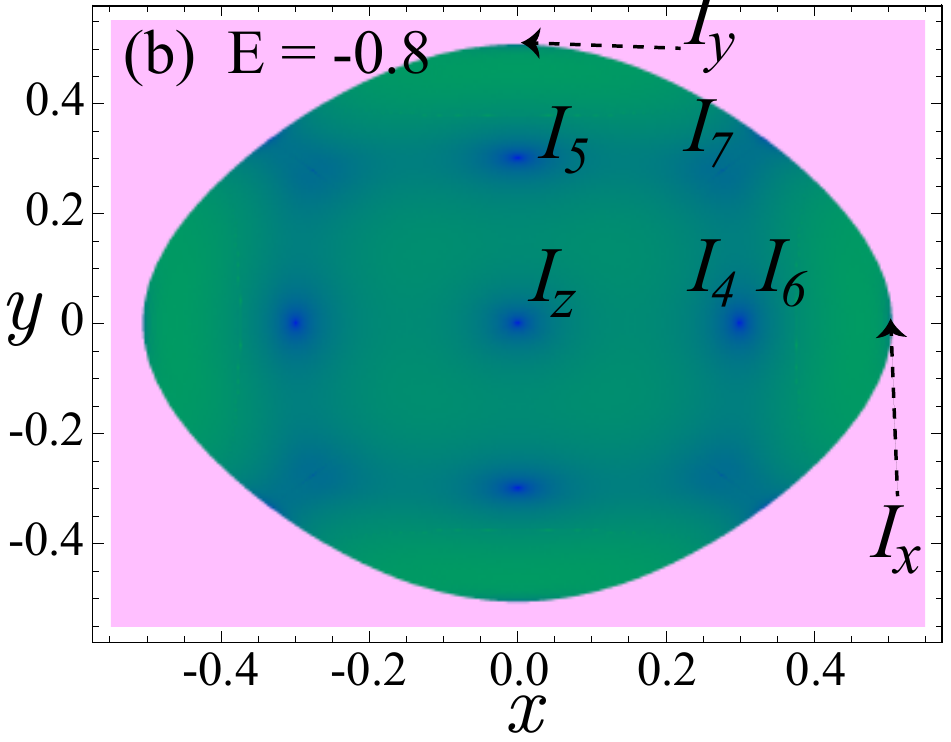}} 
\bigskip
\centerline{\includegraphics[scale=0.7]{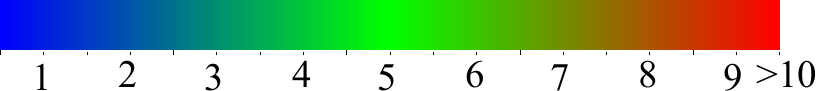}}
\caption{OFLI$^{TT}_2$ indicator color maps for  initial conditions in the plane ${\cal S}_1 =\left\{(y,z),x=v_y=v_z=0\right\}$ (upper panel) and
in the  plane ${\cal S}_2 =\left\{(x,y),z=v_x=v_y=0\right\}$ (lower panel) and  energy $E=-0.8$.
The pink color stands for energetically not allowed initial conditions.}
\label{fi:ofli2_08}
\end{figure}

If the energy of the atom is close to the potential well minimum, the phase space  is 
populated with highly regular orbits having small OFLI$^{TT}_2$ values.
This  is illustrated in~\autoref{fi:ofli2_08} by the OFLI$^{TT}_2$ maps for $E=-0.8$.
The dark-blue  points in these maps correspond to several stable periodic orbits surrounded by a set of KAM tori.
The positions of the analytical periodic orbits $I_x$, $I_y$ and $I_z$ are shown. For these OFLI$^{TT}_2$ maps, 
we have computed seven non-trivial periodic orbits,
 $I_i, \ i=1,\dots,7$, which are indicated
in the corresponding panel of~\autoref{fi:ofli2_08}.
They are also listed in Table \ref{ta:po1} with the resonance order  $m:n:k$, which means that
$w_x/w_y = m/n$ and $w_y/w_z = n/k$, with  $w_x$, $w_y$
and $w_z$ being the frequencies of each mode, and $m, n$, and $k$ integers. 

\begin{figure}[t]
\centerline{\includegraphics[scale=0.7]{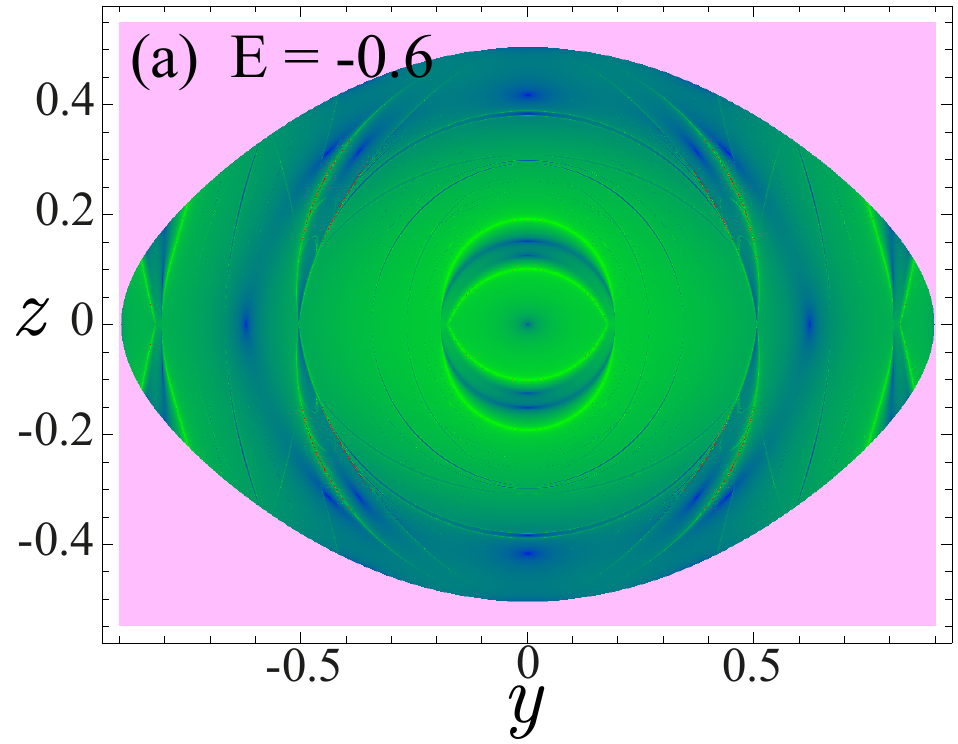} }
\centerline{\includegraphics[scale=0.7]{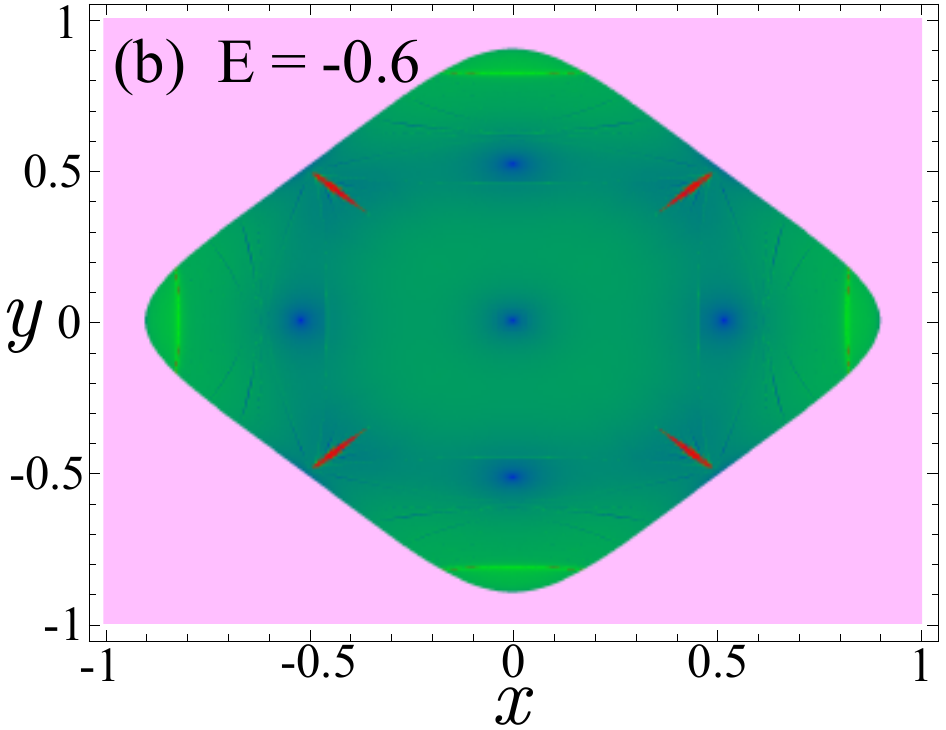}} 
\bigskip
\centerline{\includegraphics[scale=0.7]{escala}}
\caption{Same as Fig.\ref{fi:ofli2_08} but for energy $E=-0.6$.}
\label{fi:ofli2_06}
\end{figure}

\begin{figure}[t]
\includegraphics[scale=0.45]{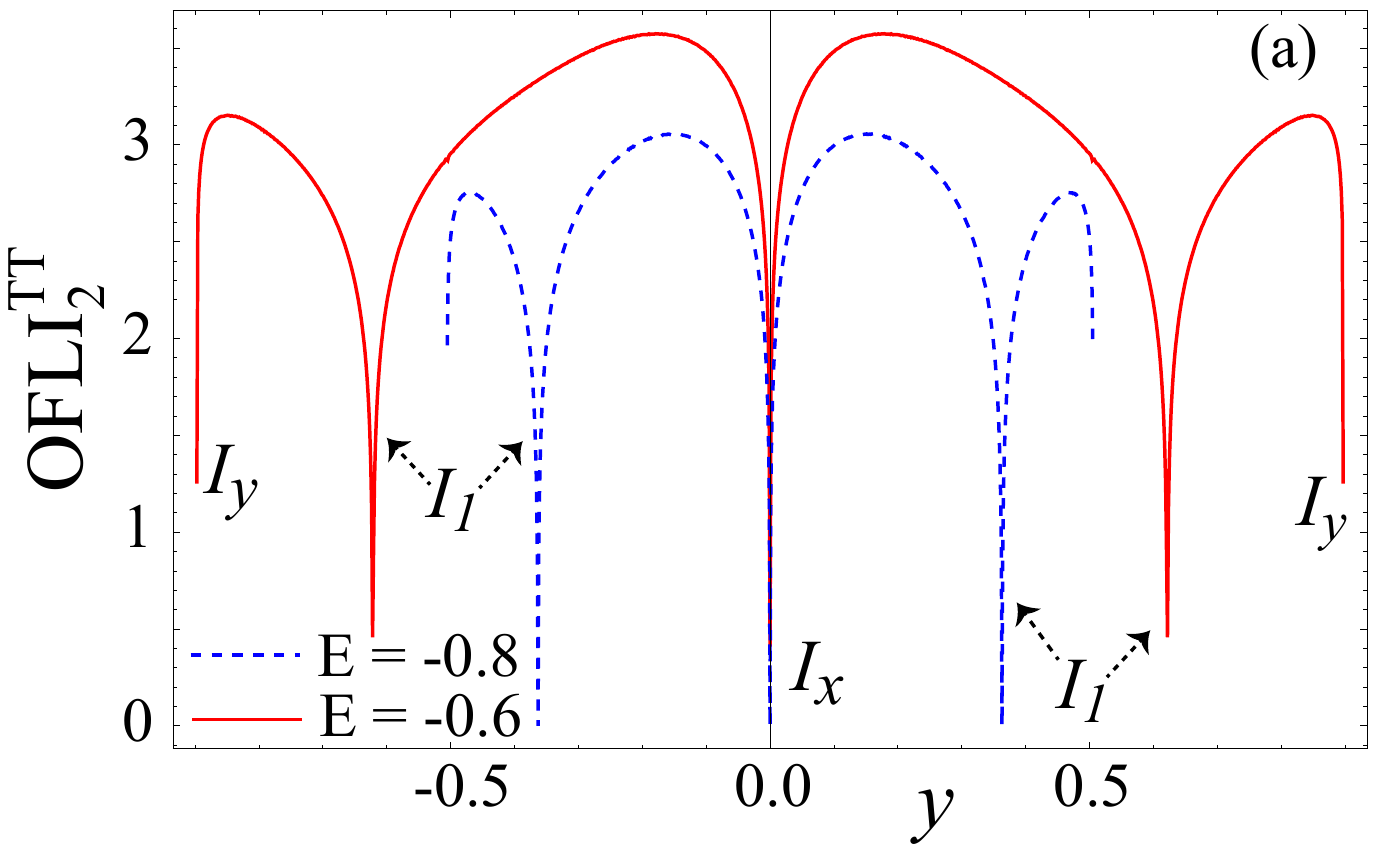}
\includegraphics[scale=0.45]{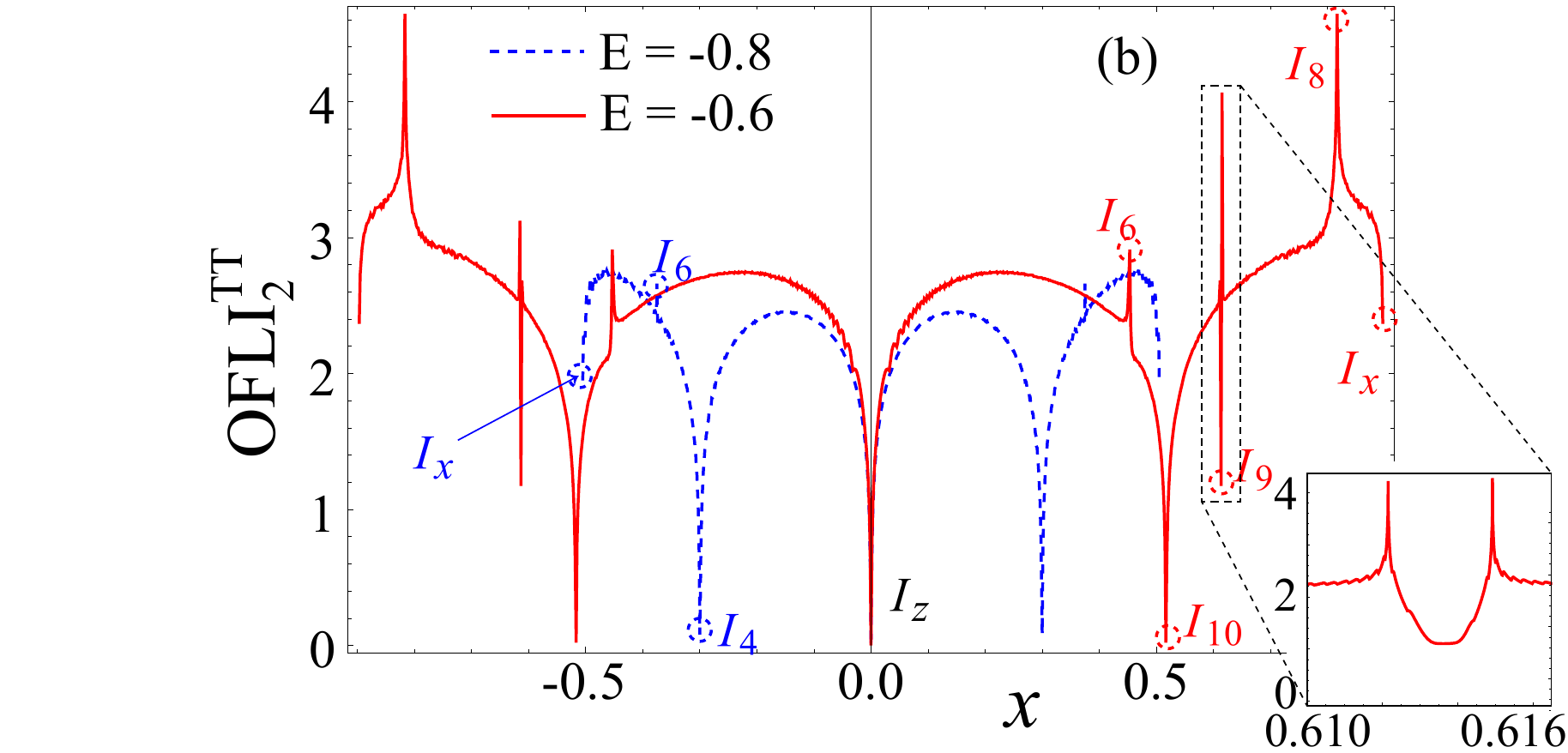}
\includegraphics[scale=0.45]{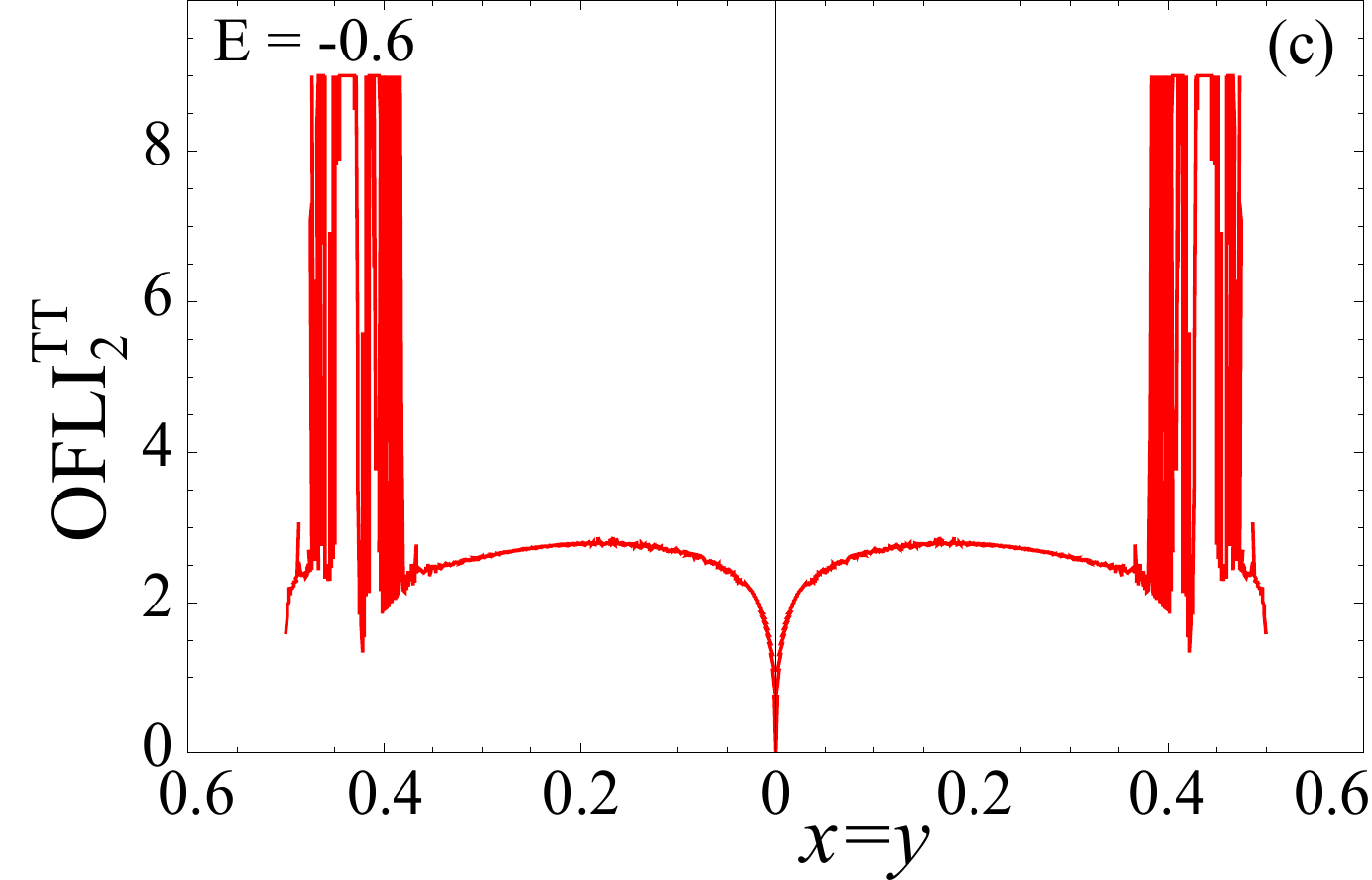}
\caption{Evolution of the OFLI$^{TT}_2$ indicator along: (a) the $z=0$ direction ($x=p_y=p_z=0$) for energies $E=-0.8$ (blue dashed line)
and $E=-0.6$ (red solid line);
(b) the $y=0$ direction ($z=p_x=p_y=0$)
for  $E=-0.8$ (blue dashed line) and $E=-0.6$ (red solid line);
(c) the $x=y$ direction ($z=p_x=p_y=0$) for $E=-0.6$.}
\label{fi:compa}
\end{figure}

For  $E=-0.6$, the phase space still presents a
predominantly regular behavior, see \autoref{fi:ofli2_06}. 
Compared to the $E=-0.8$ case, the OFLI$^{TT}_2$ maps  are dominated by
a lighter green color due to the existence of quasiperiodic orbits
with larger values of  OFLI$^{TT}_2$.
The evolution of the  OFLI$^{TT}_2$
along the $z=0$ and $y=0$ axes is presented in \autoref{fi:compa}~(a) and
\autoref{fi:compa}~(b), respectively. 
These one-dimensional plots clearly show
that larger values of OFLI$^{TT}_2$ are achieved for $E=-0.6$. Moreover, we
observe along the $z=0$ axis the major impact of the presence of the periodic orbits
$I_x$, $I_y$ and $I_1$ (see Table \ref{ta:po1}).
In \autoref{fi:compa}~(b), we detect the OFLI$^{TT}_2$ signature of
the $I_x$, $I_z$, $I_4$ and $I_6$
periodic orbits (see  \autoref{ta:po1}).
For $E=-0.6$, \autoref{fi:compa}~(b) shows the presence
of three new periodic orbits (resonances), namely $I_8$ (2:0:1),  $I_9$ (5:0:3) and  $I_{10}$ (11:0:7), which 
are also depicted in  \autoref{ta:po1}. The  dynamics is more complex for $E=-0.6$. 
For instance, the inset of \autoref{fi:compa}~(b) shows that the periodic orbit $I_9$ is
embedded between two quasiperiodic orbits with high values of OFLI$^{TT}_2$.
The phase space in \autoref{fi:ofli2_06}~(b) presents small regions of chaotic motion (in red) 
along  the direction $x=\pm y$, 
in which OFLI$^{TT}_2$  surpasses the chaotic limit  during its evolution, cf.~\autoref{fi:compa}~(c). 
Obviously, for $E<-0.5$, all orbits are bounded.
\begin{table*}[b]
\centering
 \begin{tabular}{ c c c || c c c  }
 \hline \hline\noalign{\smallskip}
Name &  Projections  & Resonance & Name &  Projections &  Resonance  \\
Type &  & $ m:n:k$ & Type &  & $ m:n:k$ \\
\noalign{\smallskip}\hline\hline\noalign{\medskip}
\parbox{0cm}{$I_1$ \\[2ex]   2D}  &  \raisebox{-0.5\height}{\includegraphics[scale=0.1]{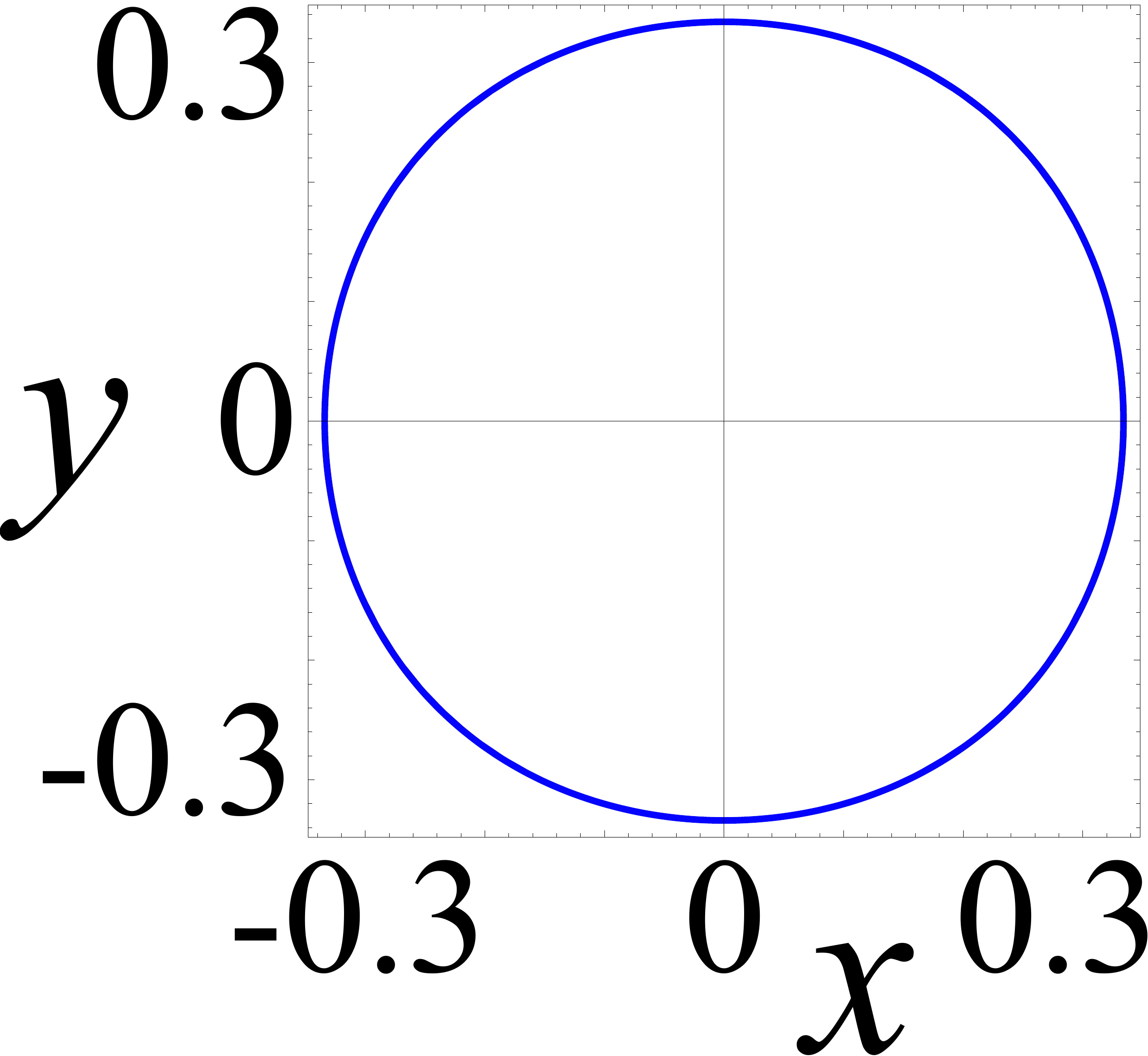}}  &   1:1:0 & \parbox{0cm}{$I_2$ \\[2ex]  2D} &   \raisebox{-0.5\height}{\includegraphics[scale=0.1]{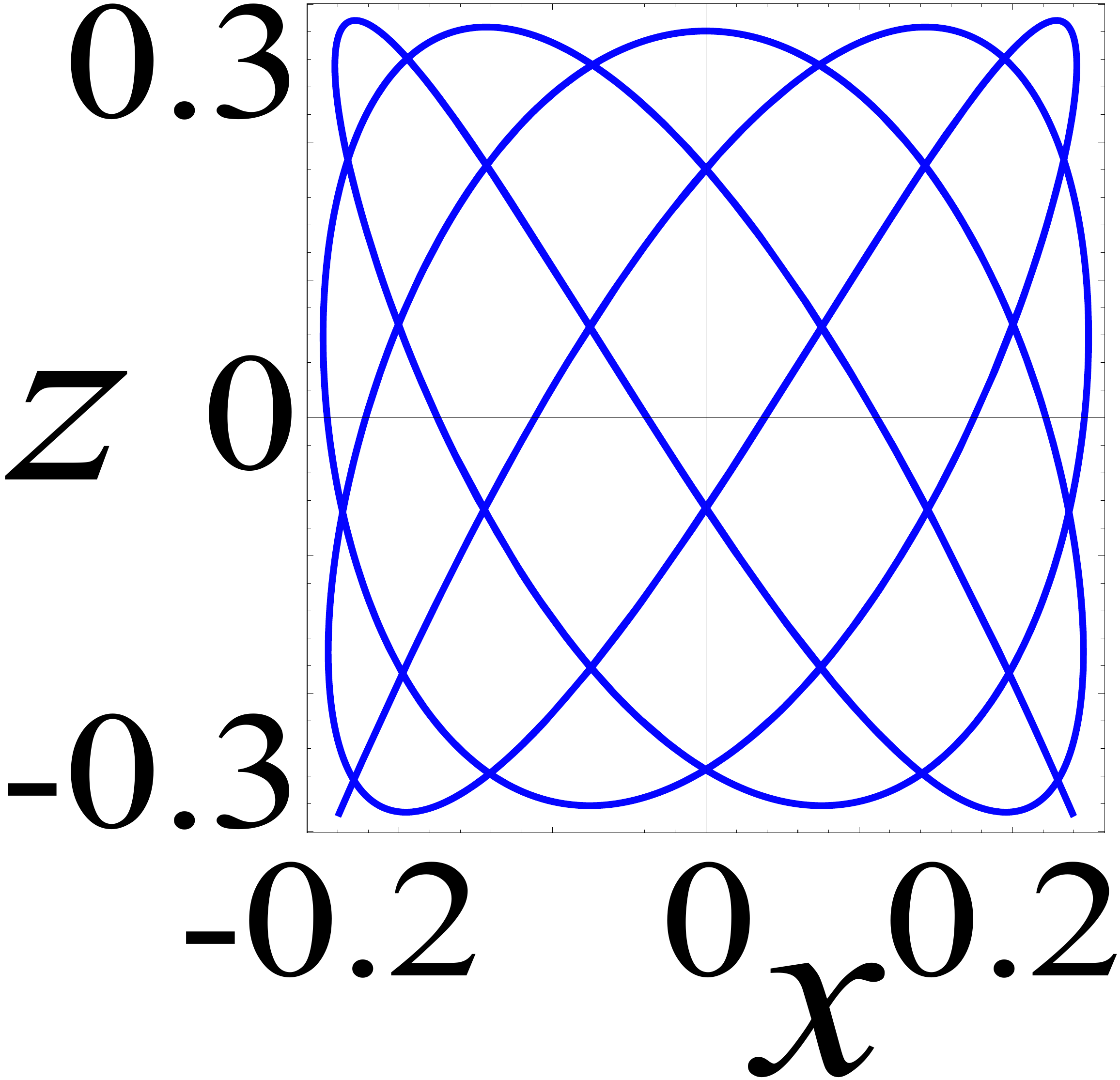}} &  10:0:7  \\   
\noalign{\medskip}
 \hline\\ 
\parbox{0cm}{$I_3$ \\[2ex] 3D}  &   \raisebox{-0.5\height}{\includegraphics[scale=0.1]{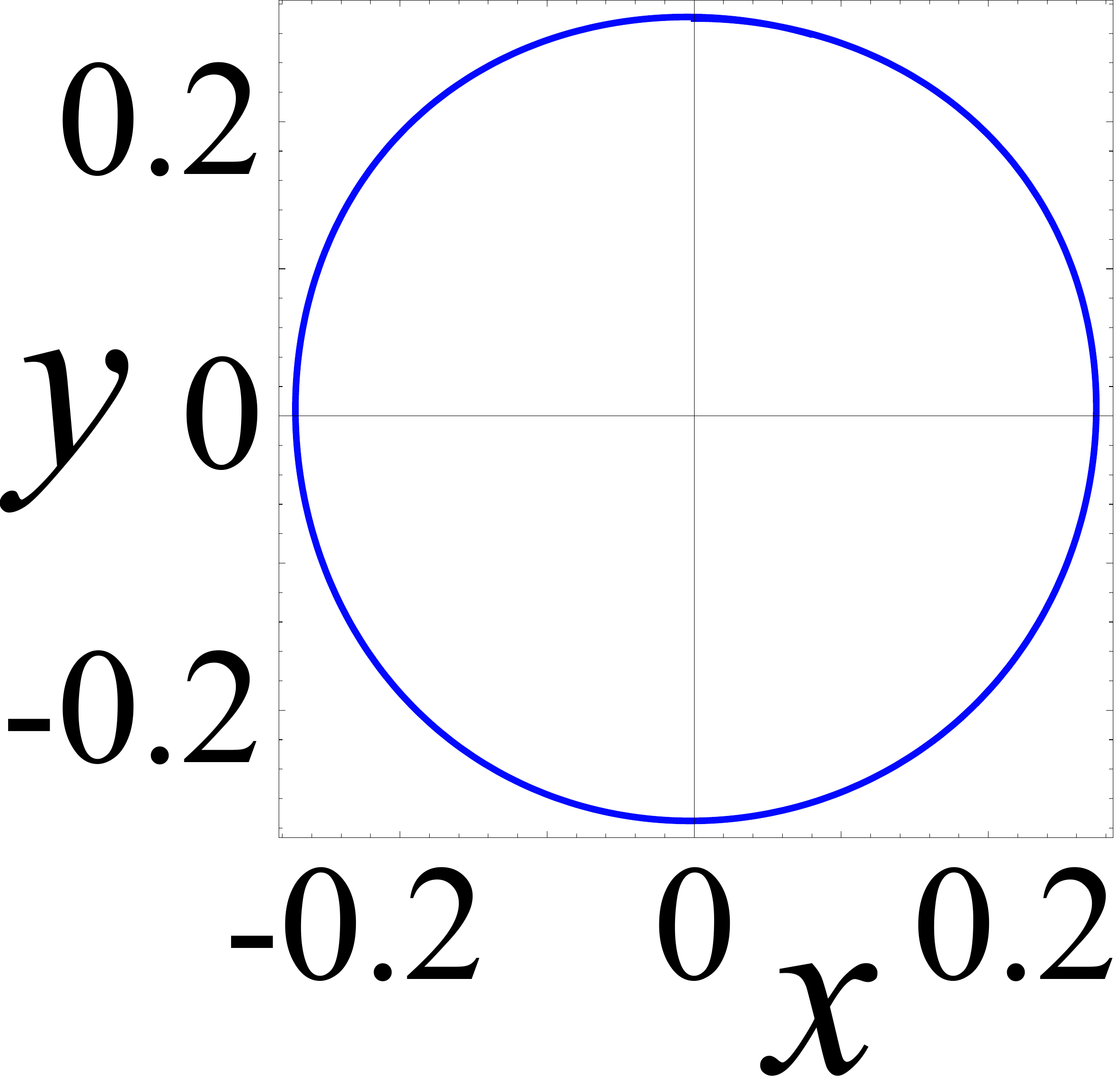}} \quad   \raisebox{-0.5\height}{\includegraphics[scale=0.1]{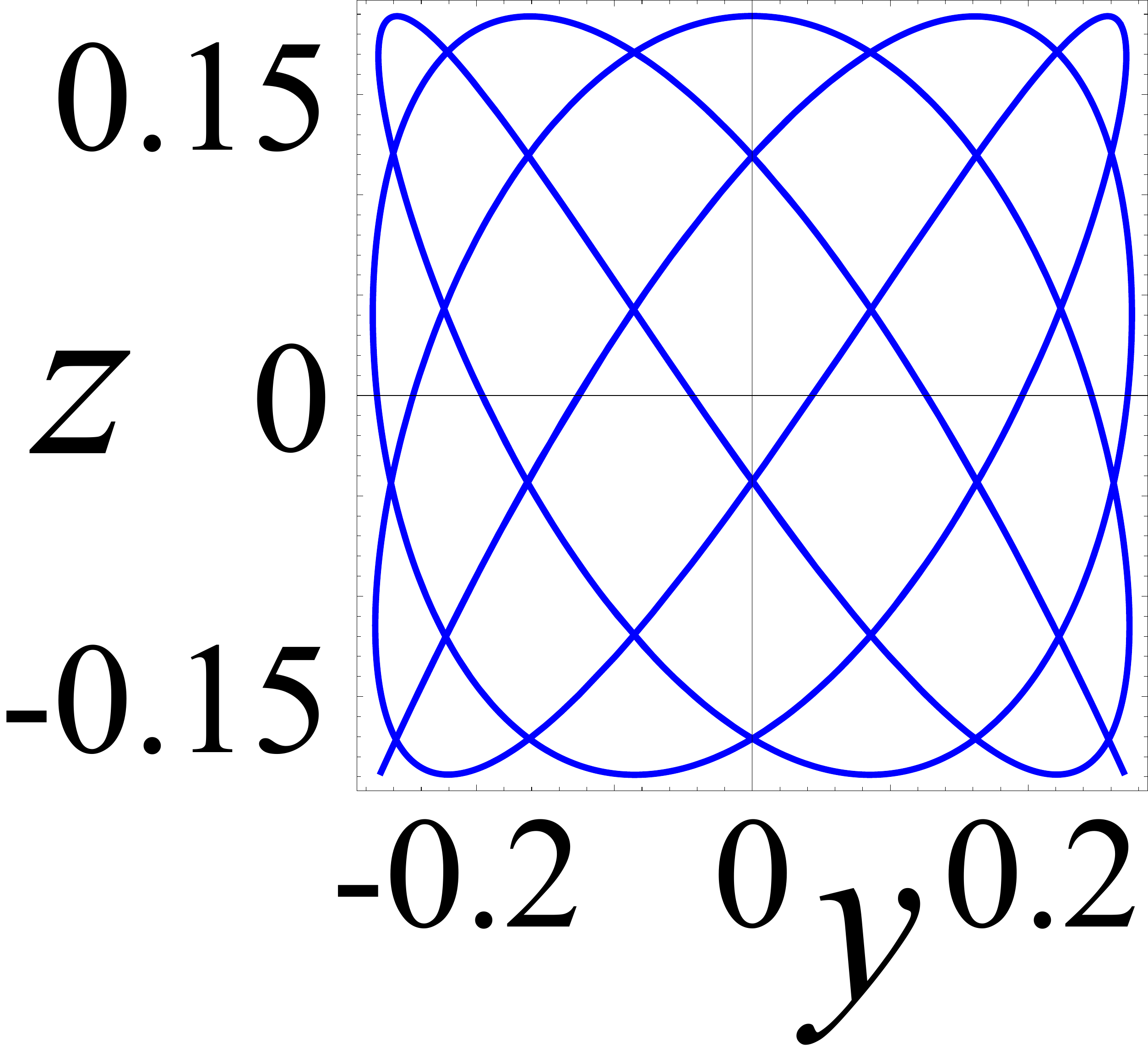}} & 10:10:7 & \parbox{0cm}{$I_4$ \\[2ex]  2D} &   \raisebox{-0.5\height}{\includegraphics[scale=0.1]{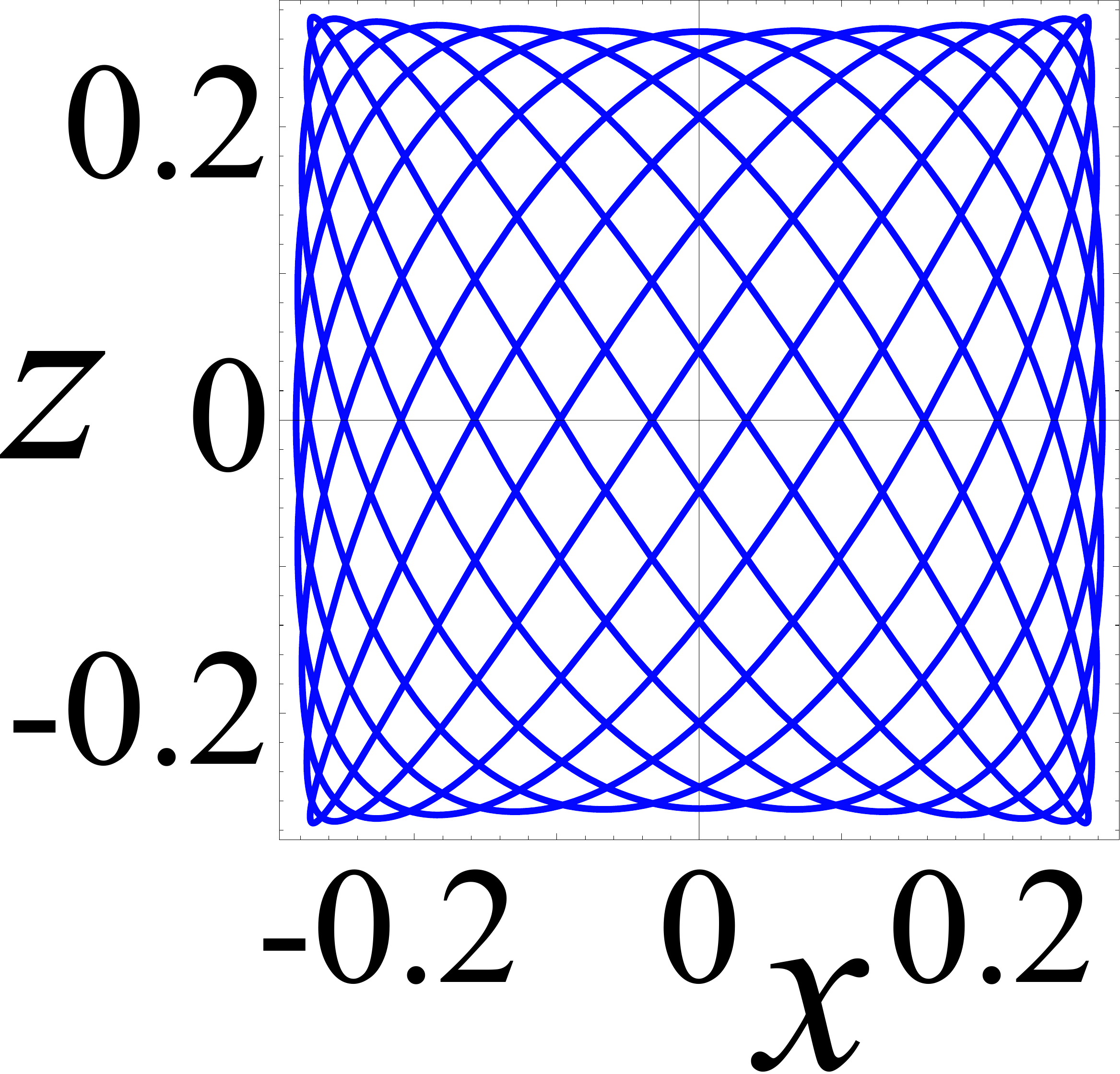}} & 13:0:9 \\
 \noalign{\medskip}
\hline\\
\parbox{0cm}{$I_5$ \\[2ex] 2D} &  \raisebox{-0.5\height}{\includegraphics[scale=0.1]{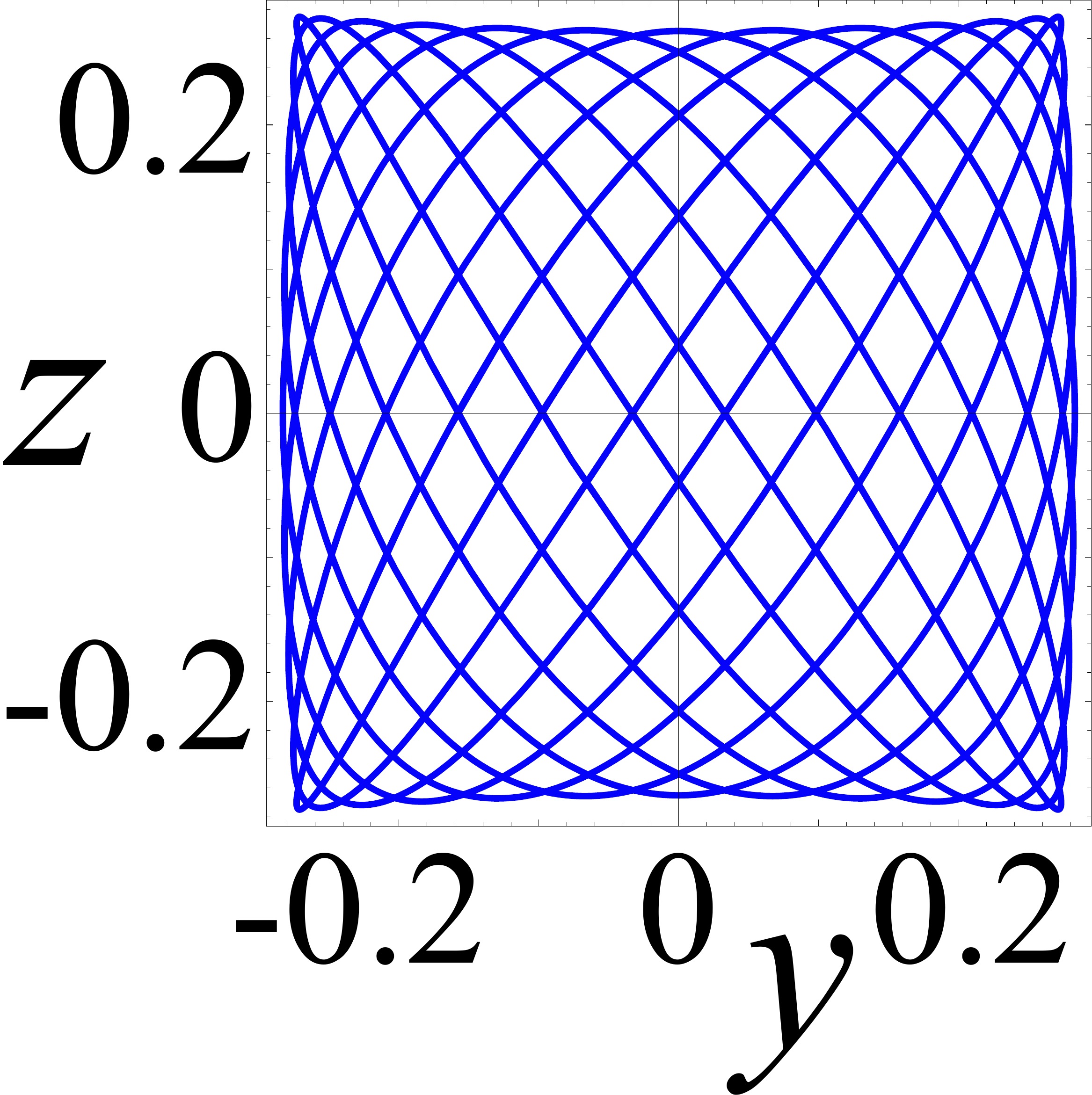}} & 0:26:17 & \parbox{0cm}{$I_6$ \\[2ex] 2D} &  \raisebox{-0.5\height}{\includegraphics[scale=0.1]{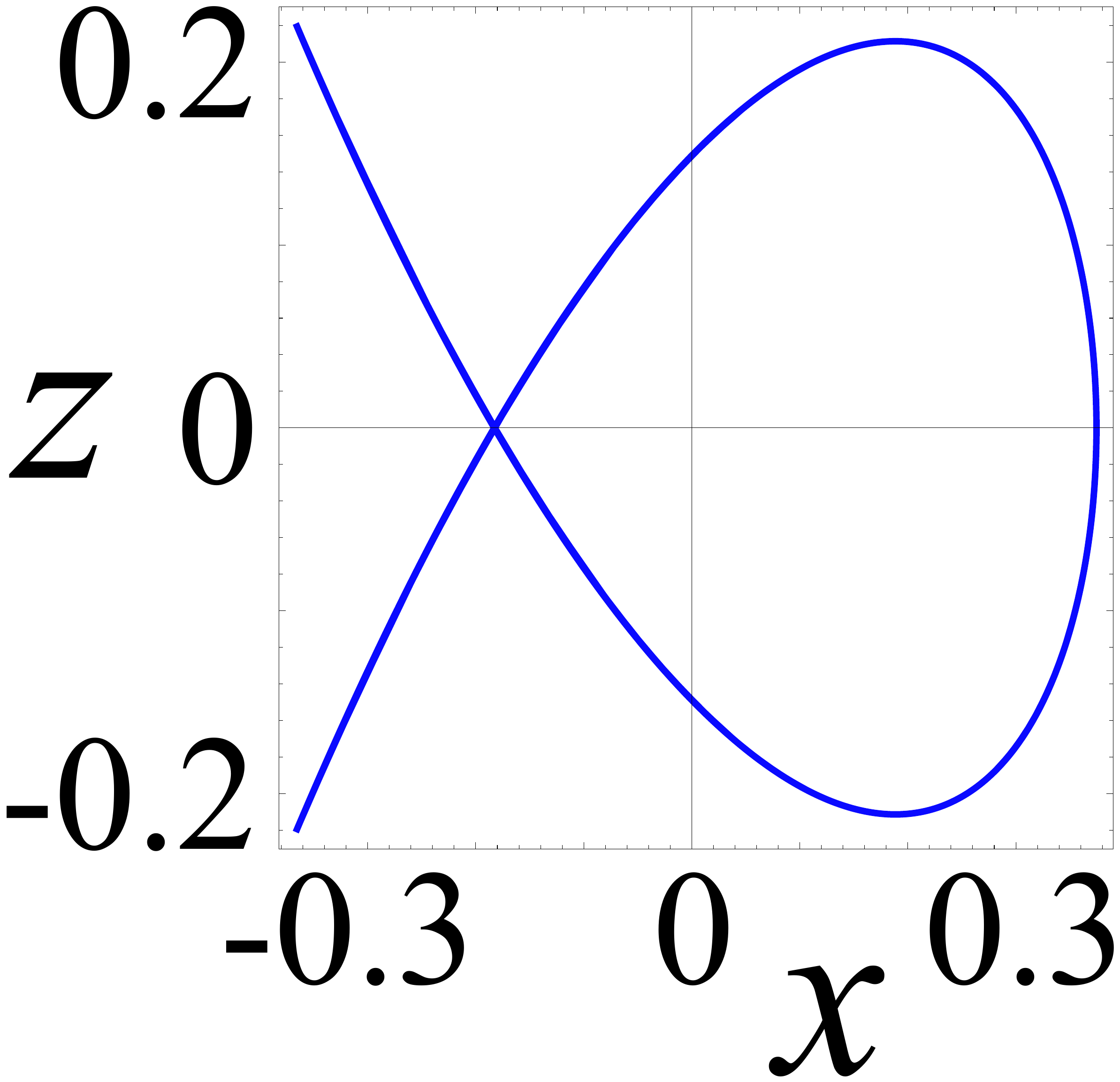}}  & 10:0:7 \\
 \noalign{\medskip}
\hline\\
\parbox{0cm}{$I_7$ \\[2ex] 3D} &  \raisebox{-0.5\height}{\includegraphics[scale=0.1]{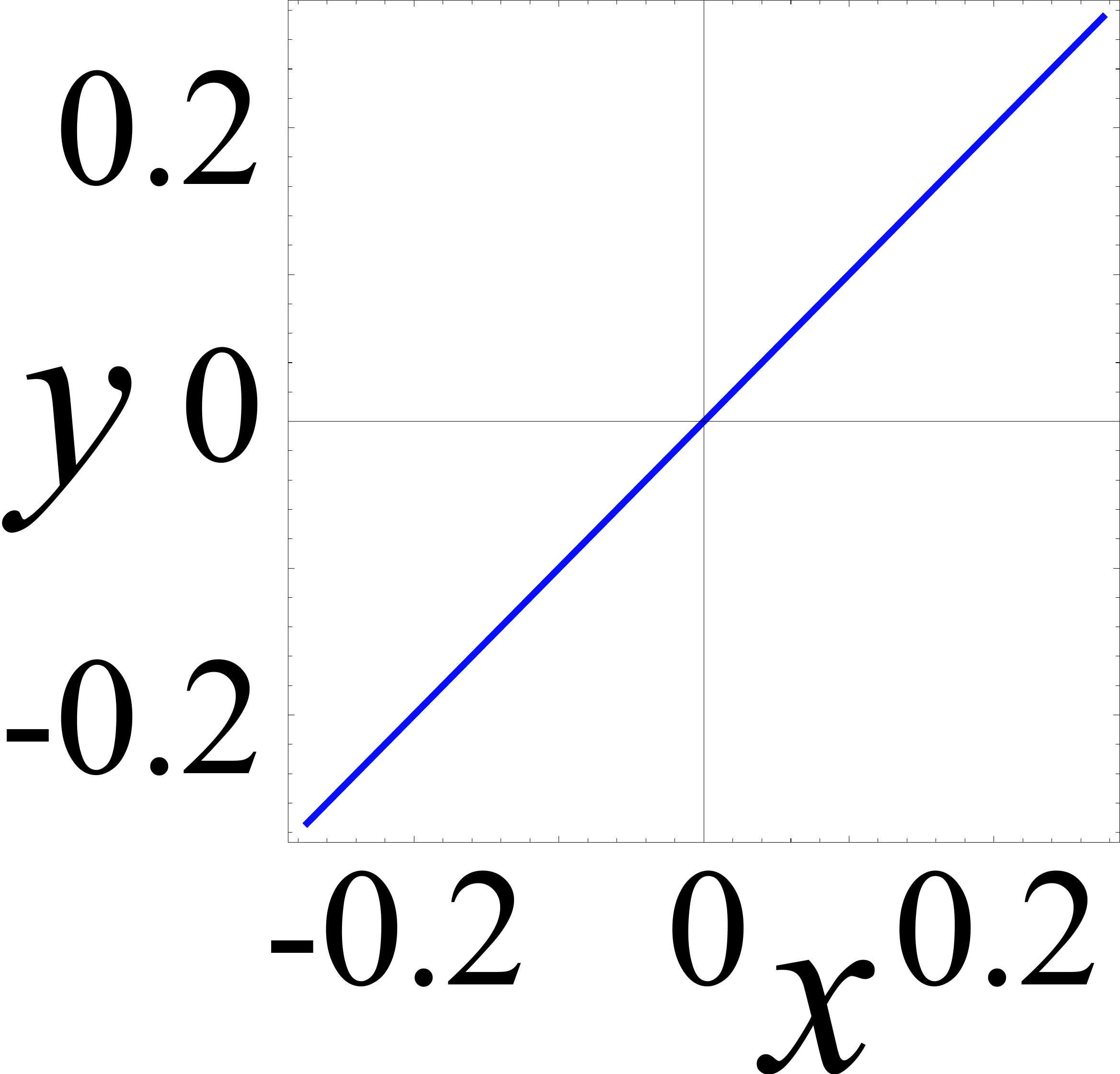}} \quad  \raisebox{-0.5\height}{\includegraphics[scale=0.1]{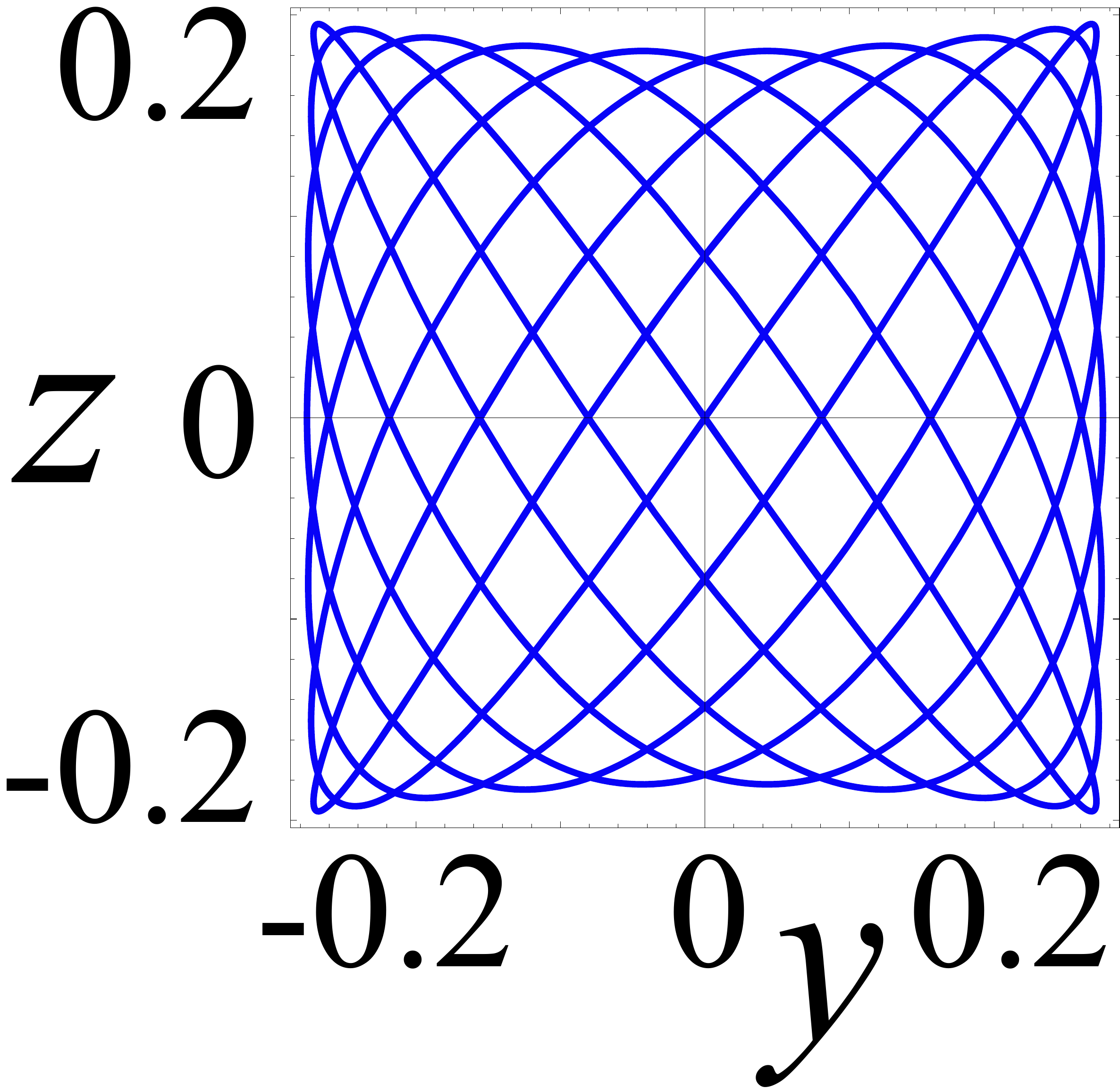}} & 7:7:5 & \parbox{0cm}{$I_8$ \\[2ex] 2D} &  \raisebox{-0.5\height}{\includegraphics[scale=0.1]{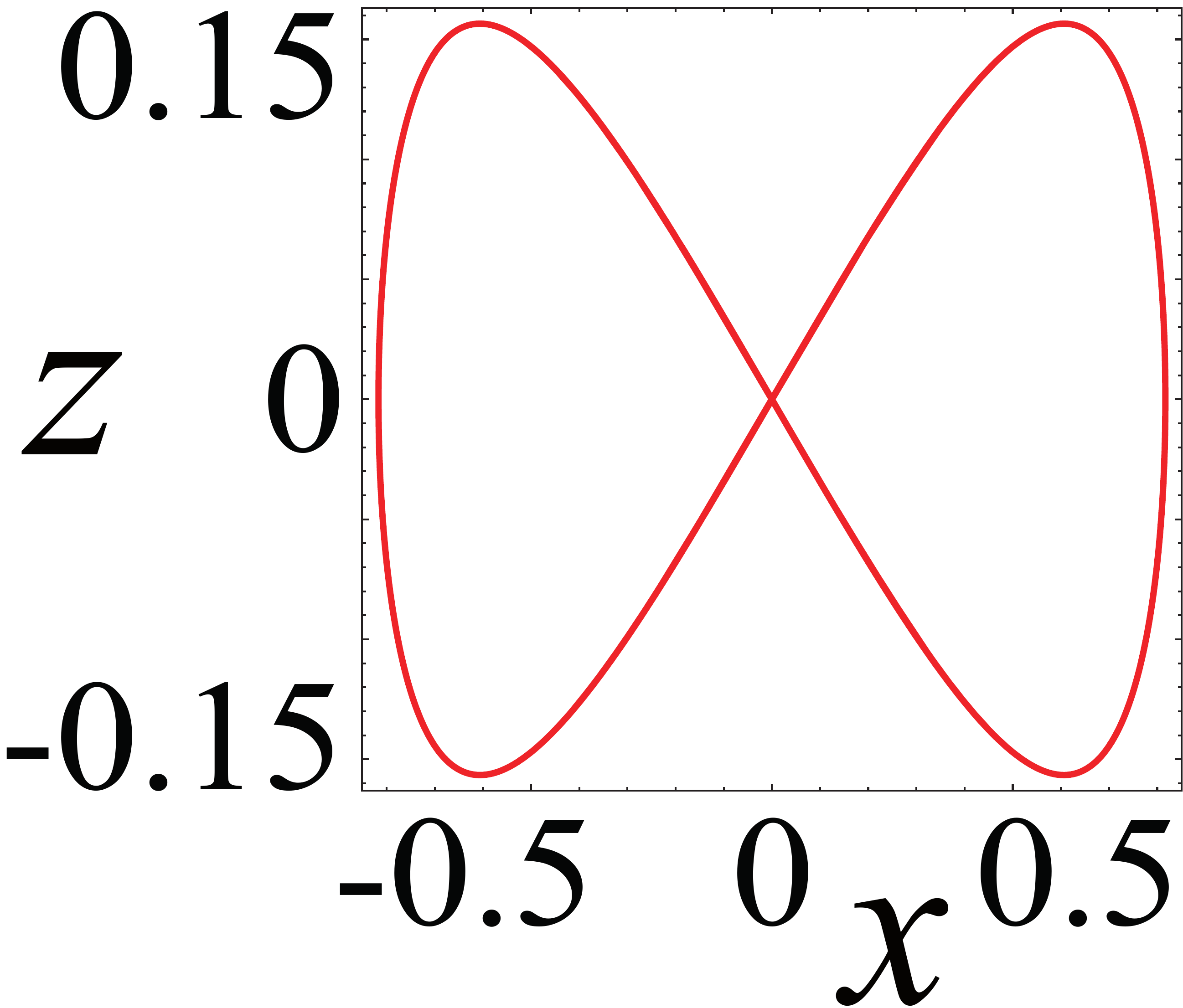}} & 2:0:1 \\
 \noalign{\medskip}
\hline\\
\parbox{0cm}{$I_9$ \\[2ex] 2D}  &  \raisebox{-0.5\height}{\includegraphics[scale=0.1]{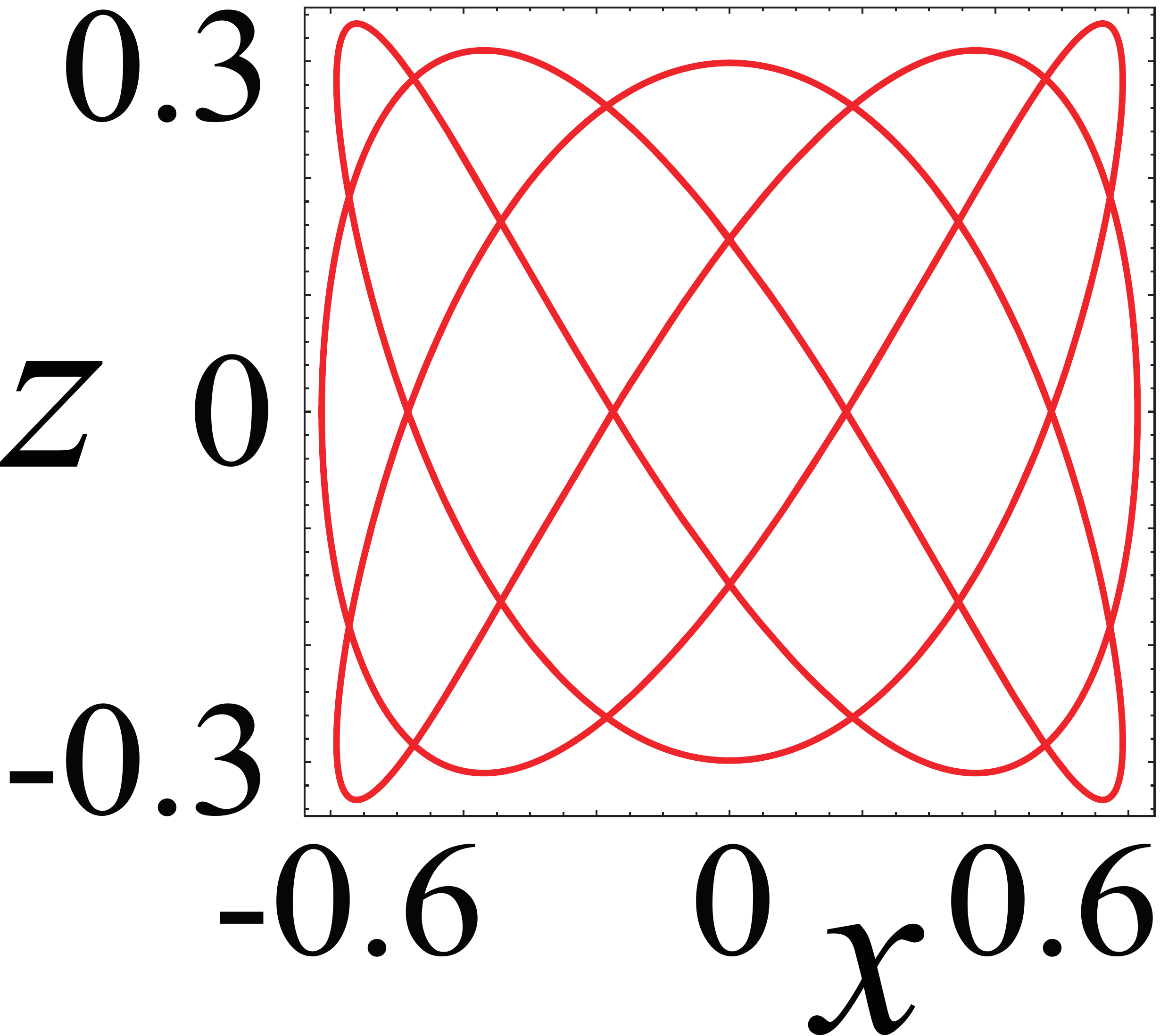}}  & 5:0:3 & \parbox{0cm}{$I_{10}$ \\[2ex] 2D} &  \raisebox{-0.5\height}{\includegraphics[scale=0.1]{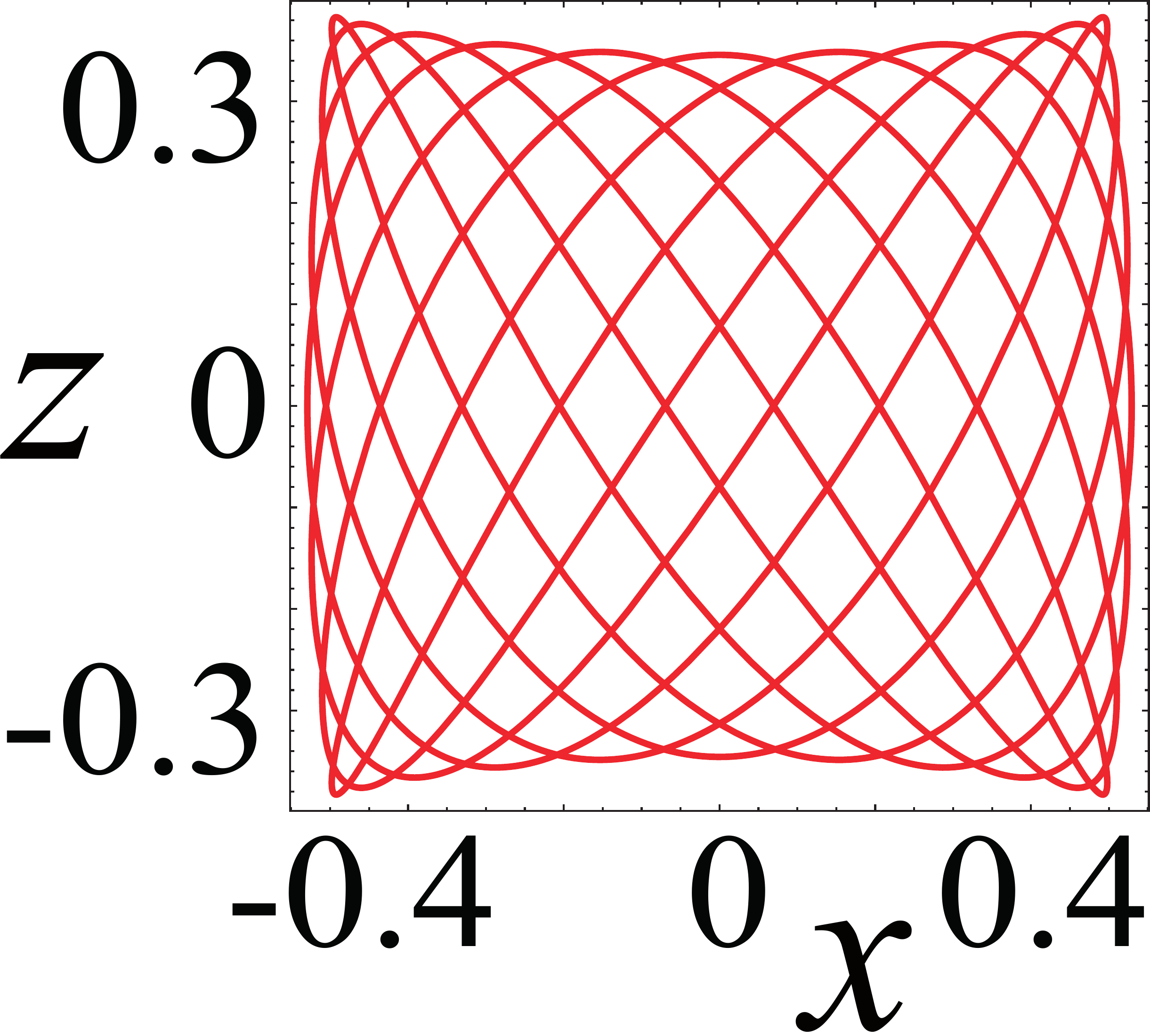}}  & 11:0:7 \\
\noalign{\medskip}\hline
\end{tabular}
%\end{tabular}}
\caption{Main non-trivial periodic orbits appearing in the OFLI$^{TT}_2$
maps in Fig.\ref{fi:ofli2_08}. These orbits are illustrated by using the 
projections on the $zx$ or $yx$ planes. We  provide the $m:n:k$ order of the resonance.
The blue and red orbits  correspond to the  energies $E=-0.8$  and $E=-0.6$, respectively.}
\label{ta:po1}
\end{table*}

%\clearpage

For $E>-0.5$, the $x$ and $y$ directions become escape channels by which
the atoms could leave the optical trap. 
Thus,  the available region of the OFLI$^{TT}_2$
map in the planar subspaces ${\cal S}_1$ and ${\cal S}_2$
are open along the $x$ and $y$ axes, respectively; whereas 
the channel along $z$ remains closed for negative energies.
Due to these escape channels, the computation of the OFLI$^{TT}_2$ indicator
is stopped if, for $t<t_f$, 
the distance of the atom  from the trap center  reaches a certain fixed threshold $r_t=5$.

\begin{figure}[t]
\centerline{\includegraphics[scale=0.7]{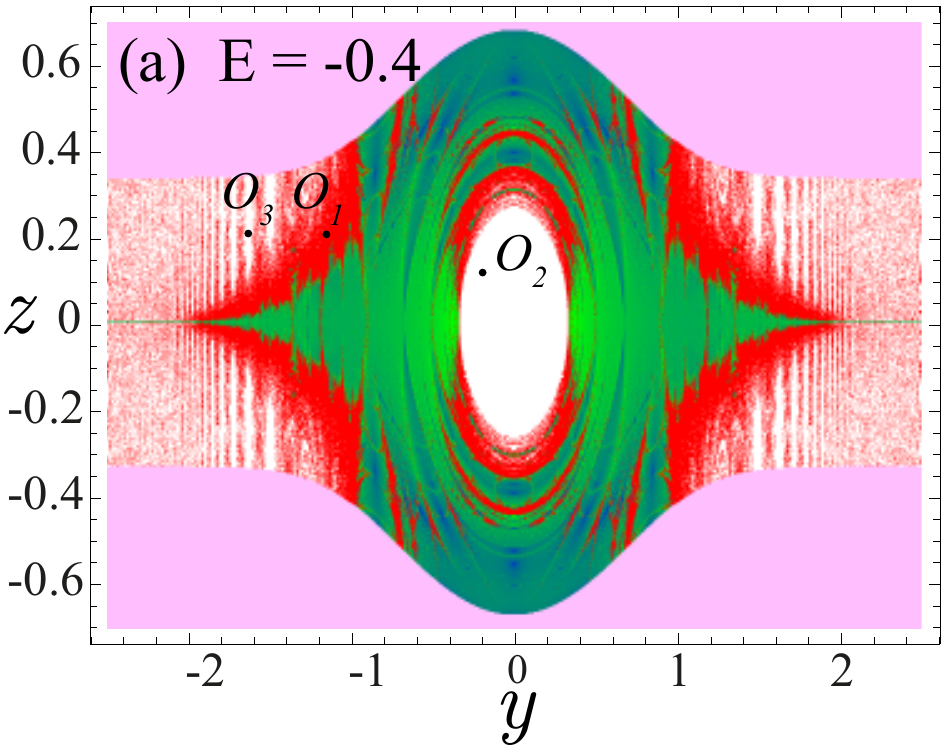} }
\centerline{ \includegraphics[scale=0.7]{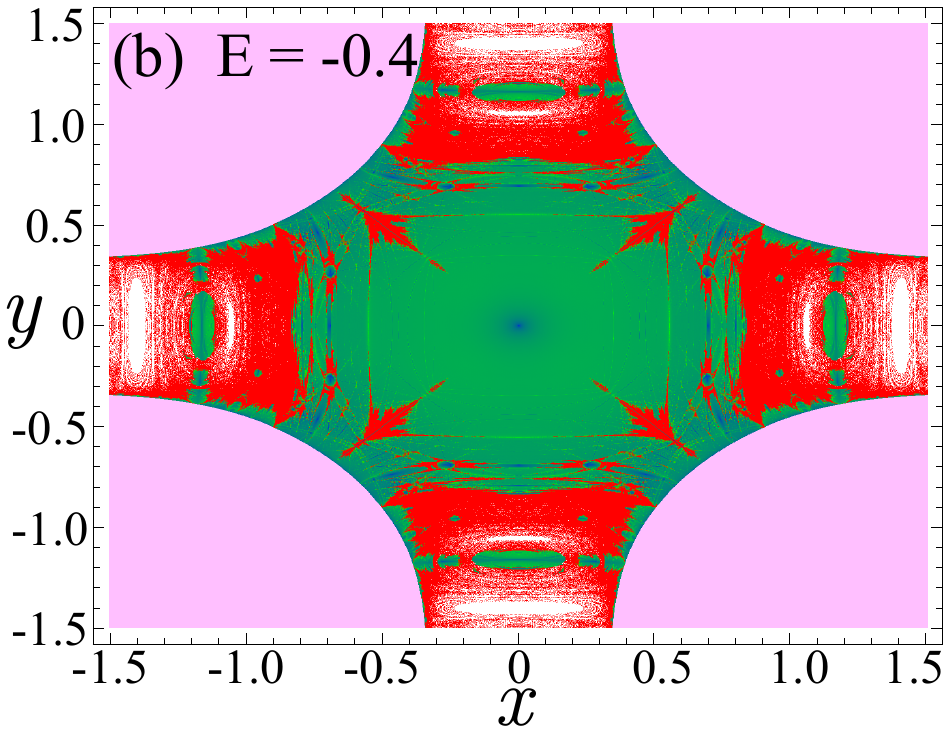}} 
\bigskip
\centerline{\includegraphics[scale=0.7]{escala}}
\caption{Same as Fig.\ref{fi:ofli2_08} but for energy $E=-0.4$.}
\label{fi:ofli2_04}
\end{figure}

\begin{figure}[h]
\centerline{\includegraphics[scale=0.45]{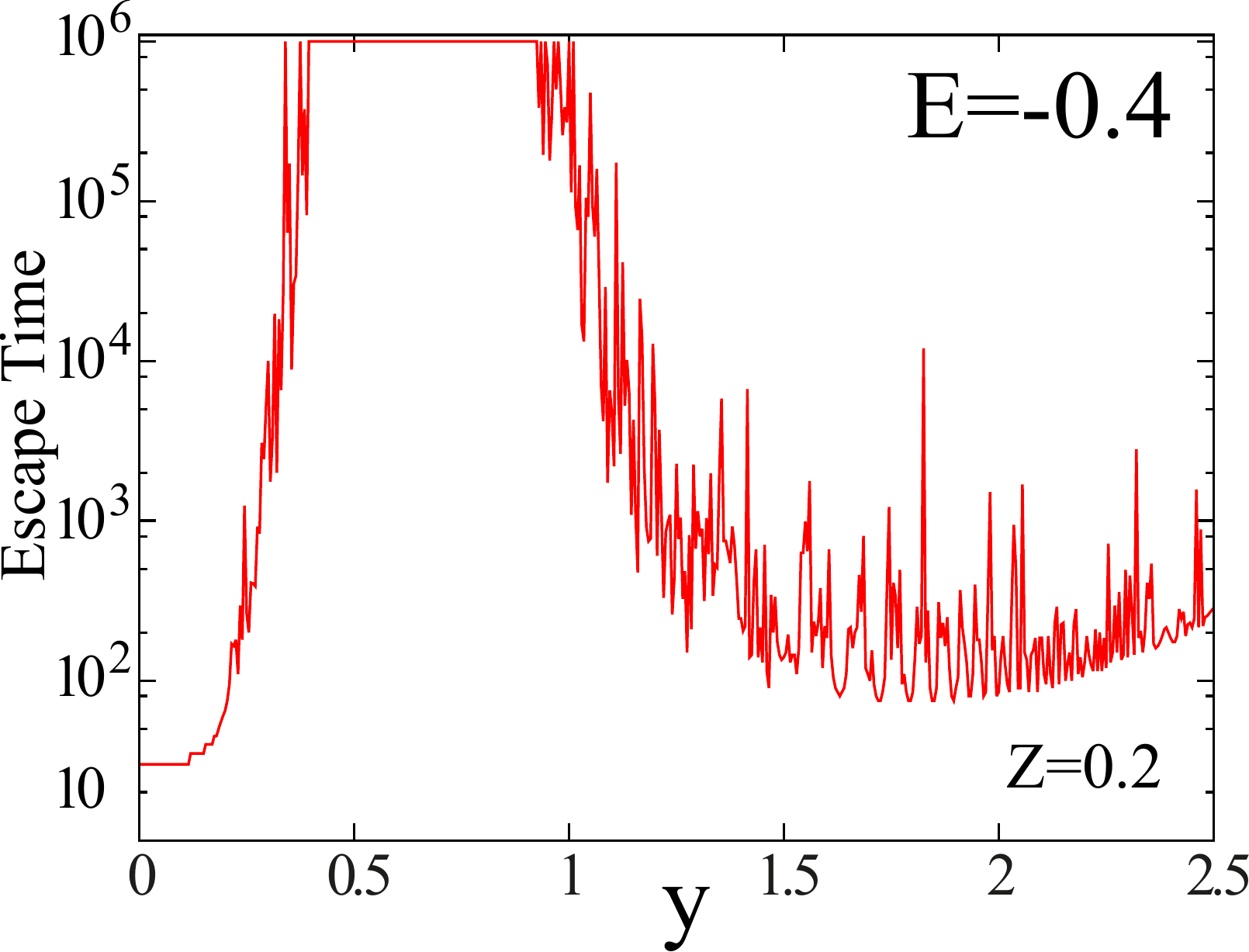} } 
\caption{Evolution of the escape time of the chaotic orbits in \autoref{fi:ofli2_04}~(a) along the $z=0.2$ direction for $E=-0.4$.
The stopping time used to obtain these chaotic orbits is $t_f=10^6$.}
\label{fi:escape_time}
\end{figure}

For $E=-0.4$,  the dynamics is more complex
 and the phase space is a mixture of chaotic and regular motion around KAM tori, see \autoref{fi:ofli2_04}. 
Note that the initial conditions in the wide red regions
of \autoref{fi:ofli2_04} give rise to chaotic orbits (in a non-strict sense)
that after being trapped for some time finally escape~\cite{footnote_mia}.
 These orbits have a transient chaotic regime while the atom remains confined in the trap. 
Once the atom leaves the trap through one of the escape channels, it becomes a free particle and its 
regular motion is  not affected by the  potential well.
In many cases, the elapsed time before the atom leaves the trap is much larger than the stopping  time $t_f=2000$
used in this work. 
This is illustrated in \autoref{fi:escape_time} with  the
evolution of the escape time of chaotic orbits in \autoref{fi:ofli2_04}(a) along the $z=0.2$ direction for $E=-0.4$. It is important to note that the stopping time used in the computation of \autoref{fi:escape_time} is  $t_f = 10^6$.

\begin{figure*}[h]
\centerline{\includegraphics[scale=0.27]{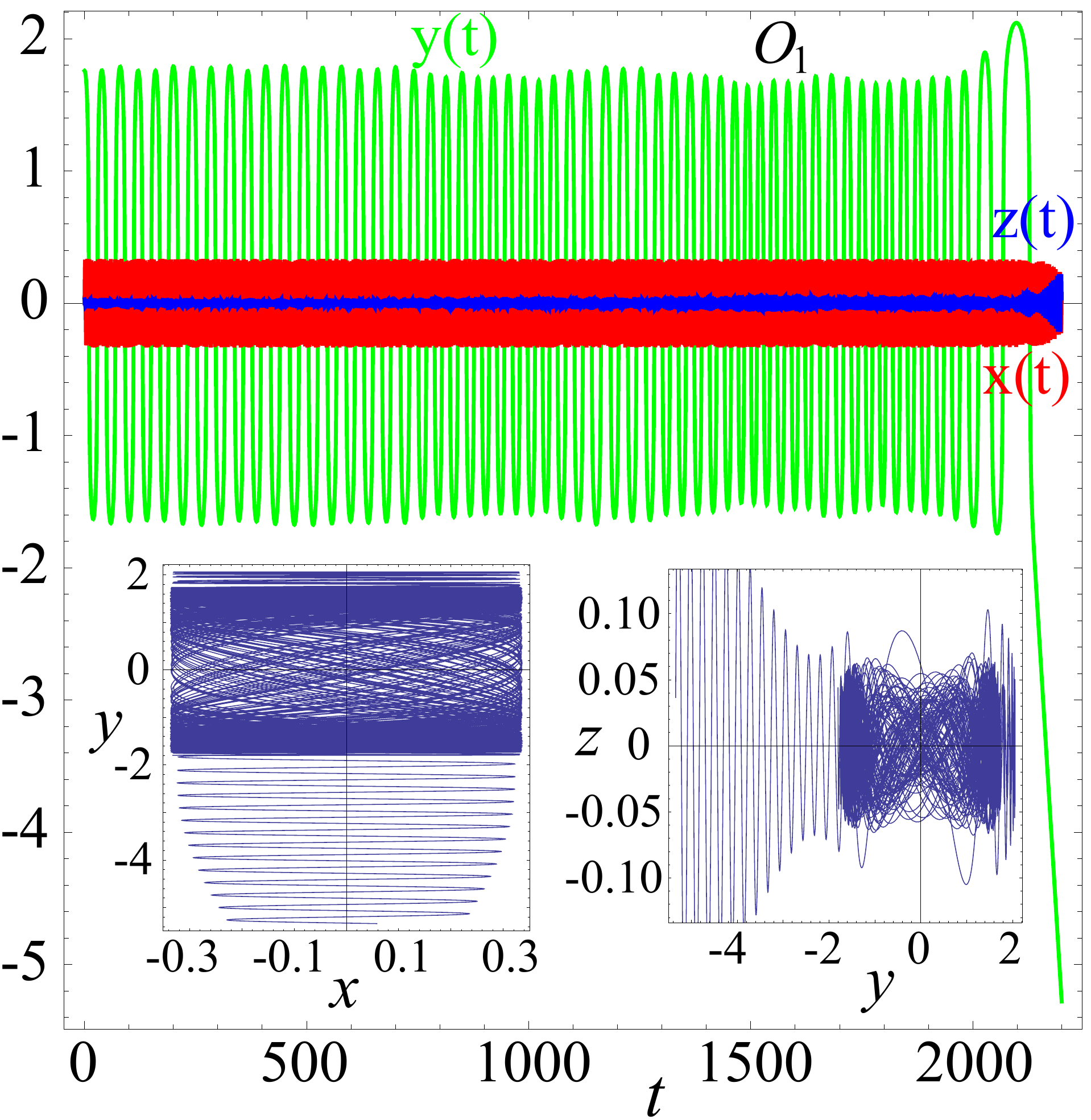} \quad 
\includegraphics[scale=0.27]{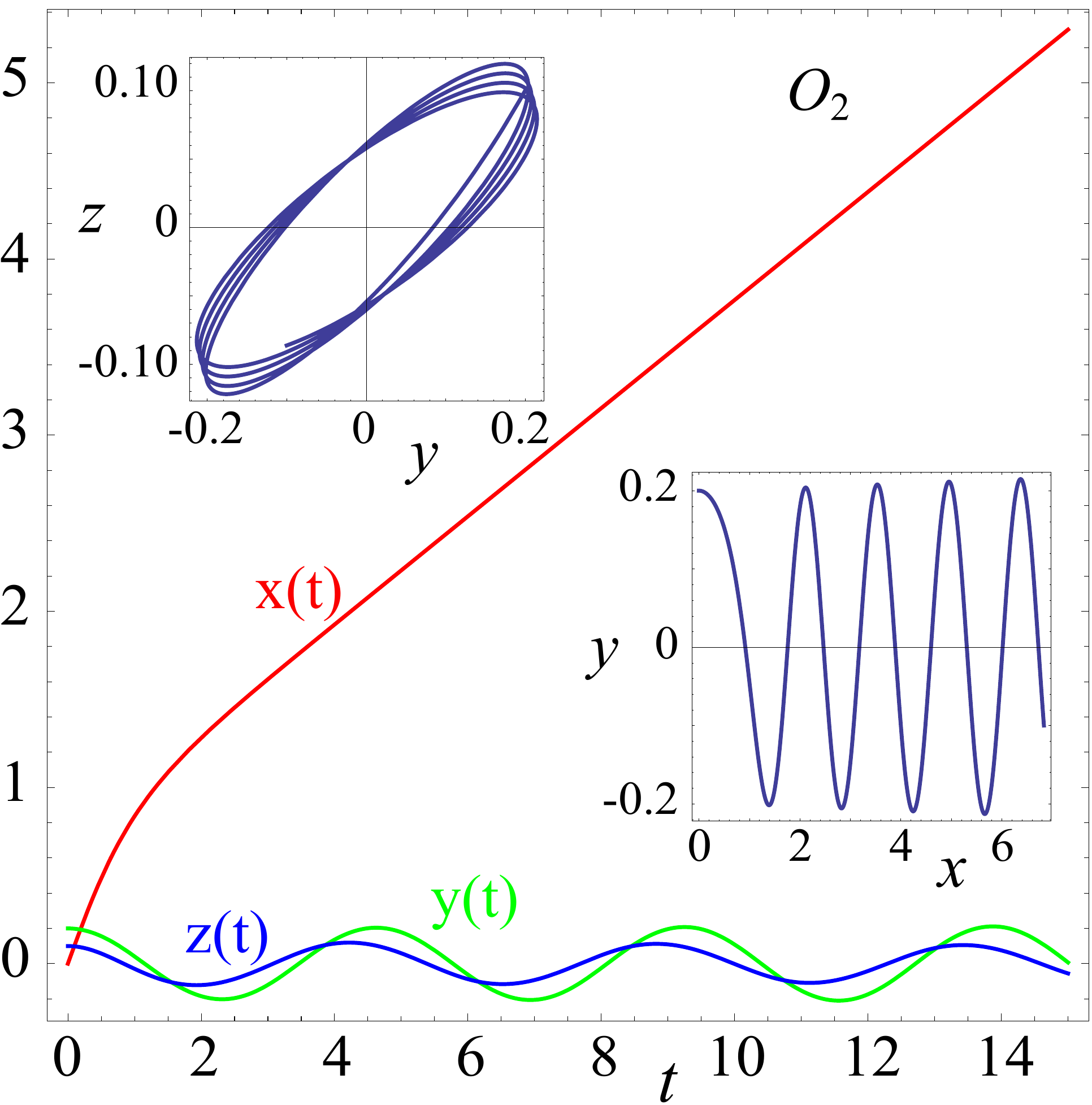} \quad \includegraphics[scale=0.27]{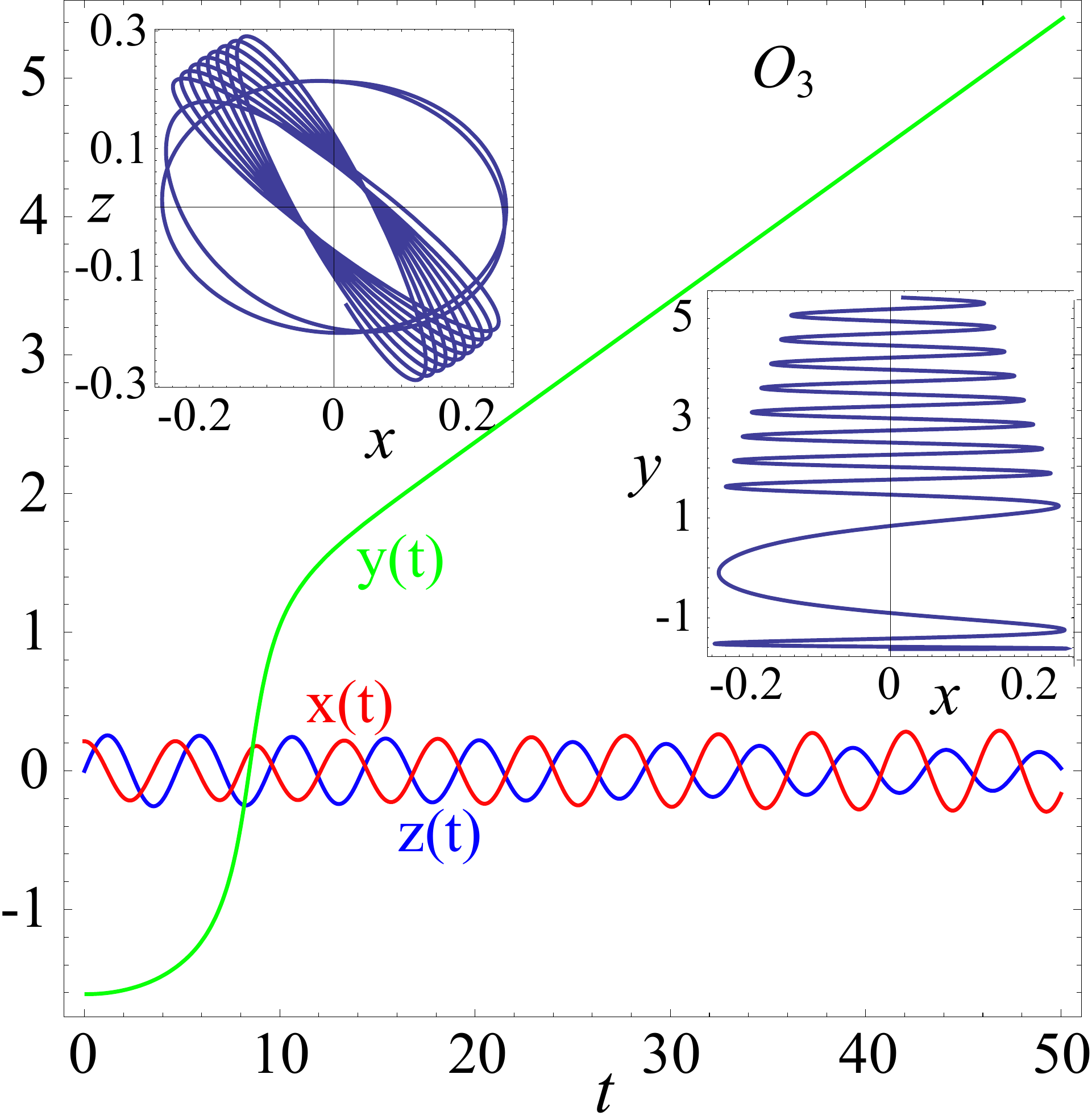}} 
\caption{Examples of the time evolution of a chaotic escape orbit ($O_1$, left panel) and two
regular escape orbits ($O_2$ and
$O_3$, central and right panels). In the insets of the panels are shown the $x-y$ and the $y-z$ projections
of each orbit.
The three orbits have the same energy $E=-0.4$, and their positions  are indicated in \autoref{fi:ofli2_04}~(a).}
\label{fi:escape1}
\end{figure*}

For more than two-degree of freedom, KAM tori do not partition phase space.
As a consequence, and due to diffusion through the Arnold's web \cite{arnold}, chaotic orbits can 
explore all the phase space 
regions not occupied by the bounded
KAM orbits,  in such a way that, sooner or later, the atoms leave the trap.
 As an example, we show  in \autoref{fi:escape1}~(a) the  long-lived escape chaotic orbit 
 $O_1$  with initial conditions  in  \autoref{fi:ofli2_04}~(a). 
 In the maps of \autoref{fi:ofli2_04} the white zones correspond to initial
conditions of orbits that escape ``quickly" from the trap, \ie, 
the elapsed time when the atom leaves the trap is much smaller than $t_f$. 
Actually,  these quick escape orbits have a regular behavior. 
In the white central region of the OFLI$^{TT}_2$ maps in  \autoref{fi:ofli2_04}~(a), 
the regular trajectories  escape along the $x$-axis, such as, \eg,
the orbit  
$O_2$ plotted in \autoref{fi:escape1}~(b). % and with  initial conditions shown in \autoref{fi:ofli2_04}~(a). 
Whereas, in the regular white areas surrounded by chaotic motion in the 
left and right sides of  \autoref{fi:ofli2_04}~(a), the atoms
quickly leave the trap along the $y$ axis, \eg,  the orbit $O_3$ depicted in \autoref{fi:escape1}~(c). 
The central region  of the OFLI$^{TT}_2$ maps in  \autoref{fi:ofli2_04}~(b) 
corresponds to the region around the $I_z$ rectilinear periodic orbit, which  has no accessible escape channel for negative 
energies.  In the right and left (upper and lower) white areas
embedded between chaotic motion regions in \autoref{fi:ofli2_04}~(b)
there are regular trajectories  that quickly leave the trap along the $x$ ($y$) -axis,
these orbits are  similar to those presented in \autoref{fi:escape1}. 
The orbits with initial conditions in the white regions of \autoref{fi:ofli2_04}~(a) and (b) have confinement times in the trap 
being very short compared 
to the orbits within the red regions. It is worth noticing that, for a fixed 
energy  $E$, the size of these white regions do not increase if the integration stopping time $t_f$  is increased.

\begin{figure*}[t]
\centerline{
\includegraphics[scale=0.55]{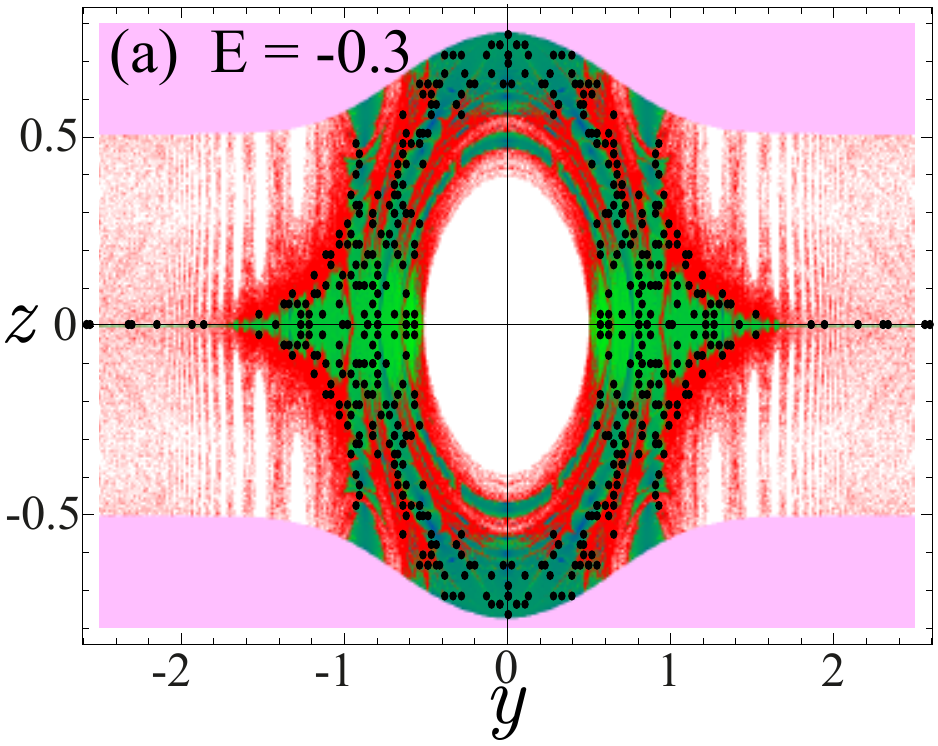} 
\includegraphics[scale=0.55]{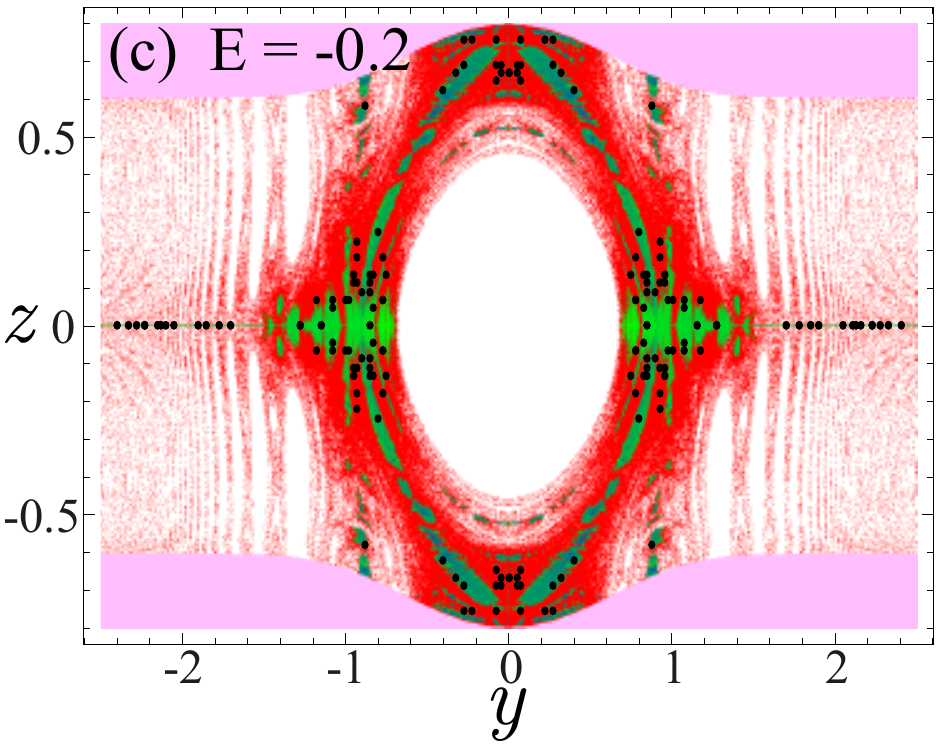} 
\includegraphics[scale=0.55]{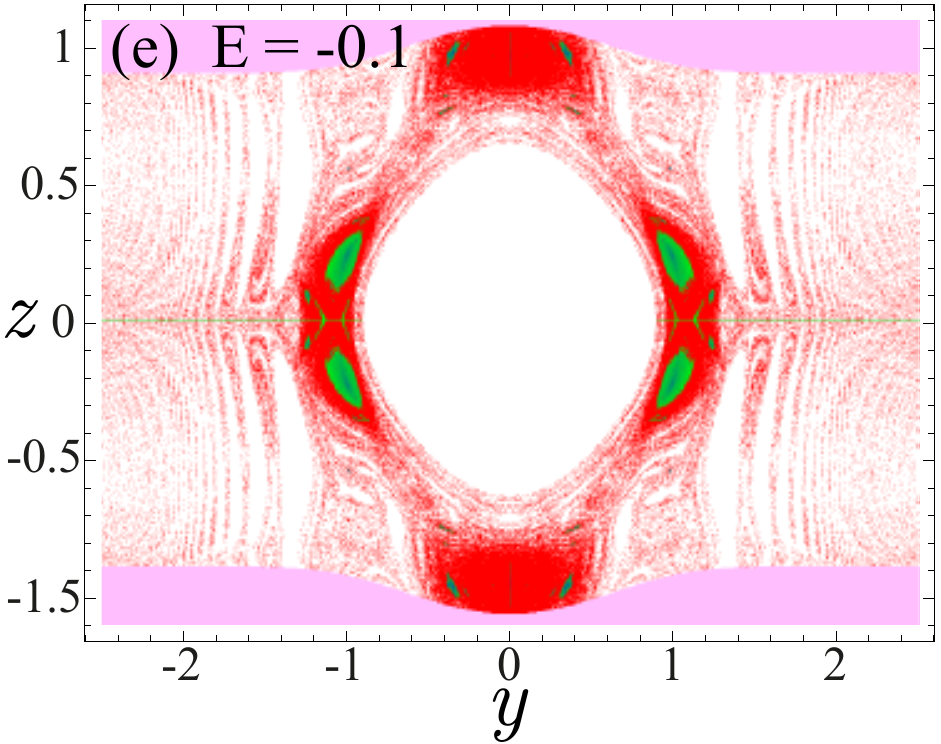}}
\centerline{
\includegraphics[scale=0.55]{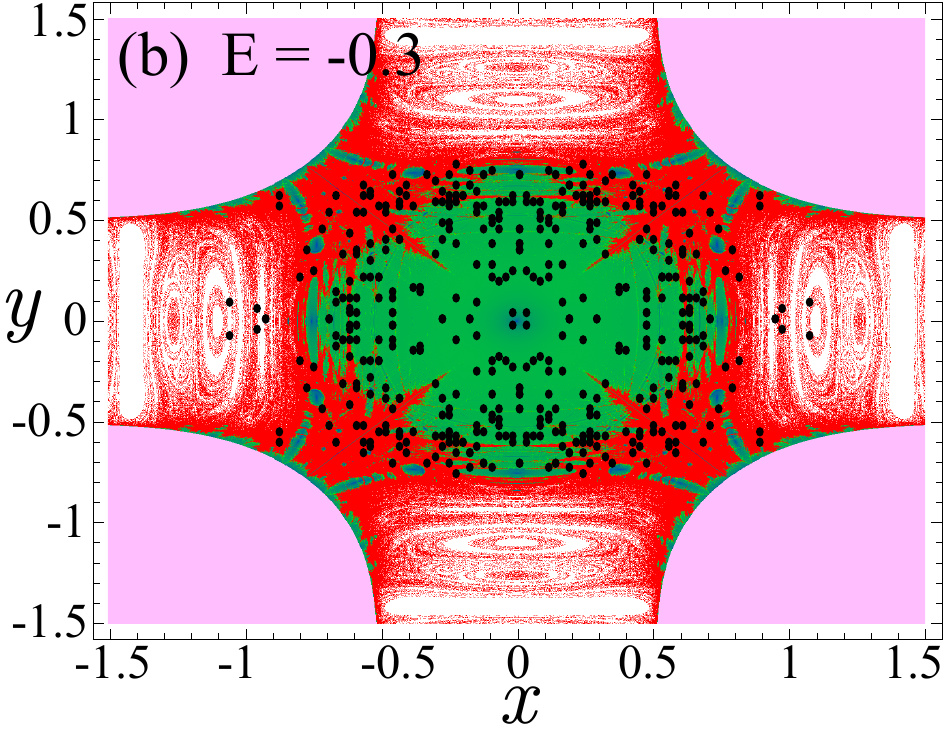} 
\includegraphics[scale=0.55]{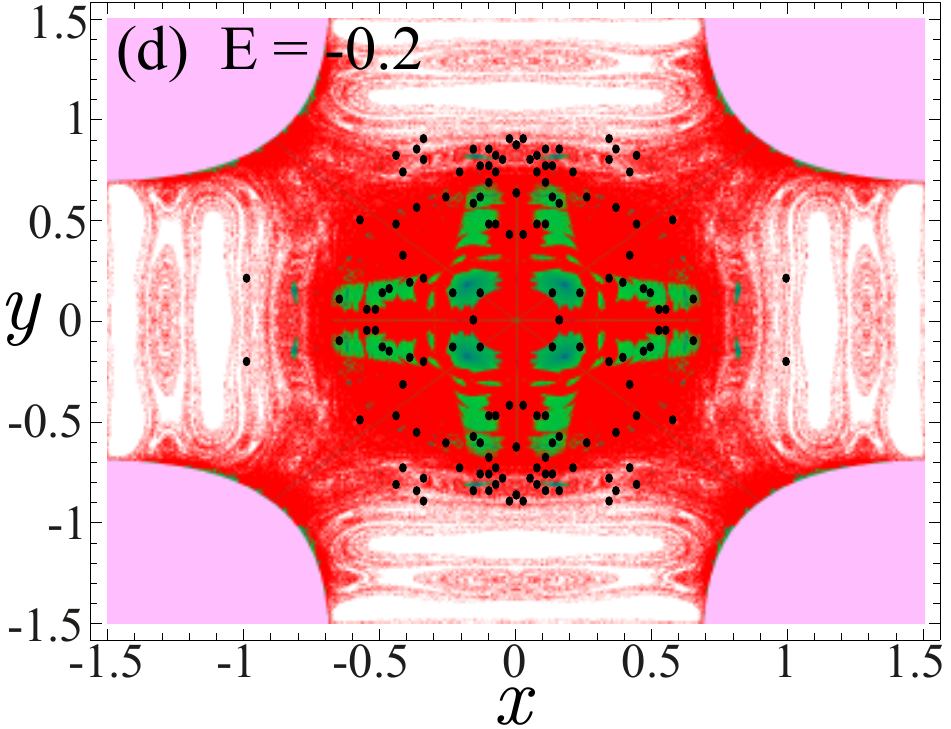} 
 \includegraphics[scale=0.55]{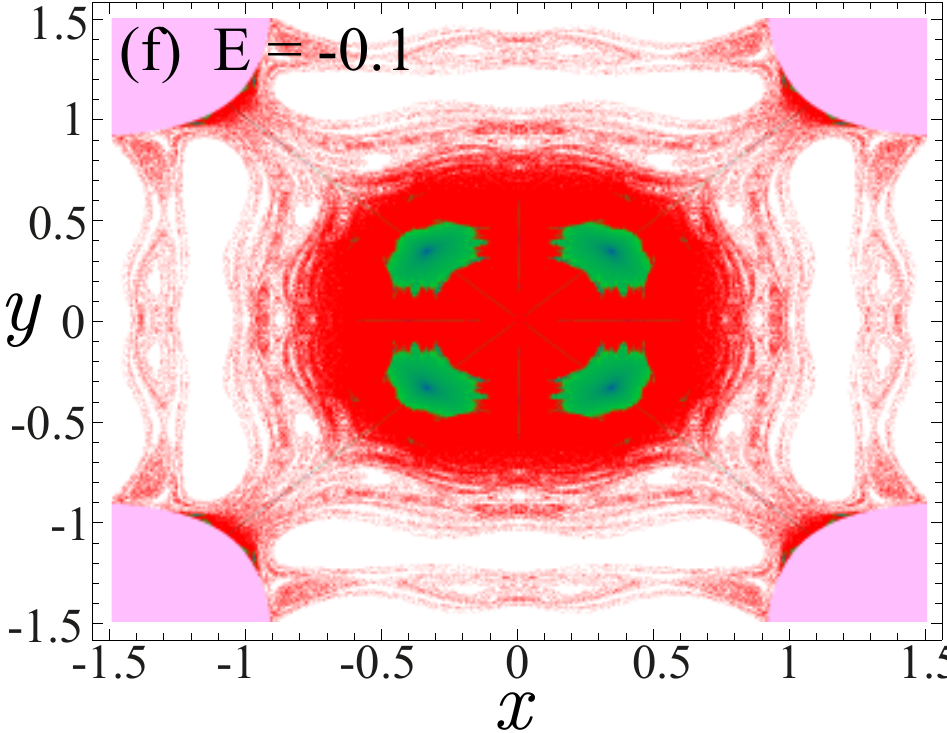}}
\bigskip
\centerline{\includegraphics[scale=1]{escala}}
\caption{Same as Fig.\ref{fi:ofli2_08} but for energies $E=-0.3$ (a) and (b) panels,
$E=-0.2$ (c) and (d) panels and $E=-0.1$ (e) and (f) panels.
The black points in panels (a), (b), (c) and (d) indicate the initial conditions of the atomic beam propagating along
the $x$ and $z$ axes which are confined when the optical trap is switched on, see Sec. \ref{sec:trap2d} for more details.}
\label{fi:ofli2_030201}
\end{figure*}

By further increasing the atomic energy,  the size of the central gap increases because the  number of orbits having 
access to the $x$-axis escape channels is enhanced, see \autoref{fi:ofli2_030201}~(a), (c) and (e) for 
$E=-0.3$, $-0.2$ and $-0.1$, respectively. 
Analogously, the escape regions along the $x$ and $y$ axes become also larger. 
At the same time, the red chaotic escape regions increase in size whereas the green regular 
bounded regions shrink, see \autoref{fi:ofli2_030201}~(b), (d) and (f).
Indeed, bounded regular motion can only be found in small regions entirely
surrounded by a chaotic sea. However, even for $E=-0.1$, where most of the orbits are unbounded 
(whether regular or chaotic), it is still possible to find regions of trapped regular motion.

\section{Classical sudden trapping mechanism}
\label{sec:trap2d}
In the cold (and ultracold) regime, 
the availability of confined atomic ensembles provides major advantages as compared to atomic beams. 
The trapped ensemble allows to perform a wealth of
interesting controlled experiments such as  
high-precision spectroscopy or single atom manipulation~\cite{PhysRevA.67.033403}.
Experimentally,  the trapping of atoms requires the manipulation of  their collective motion by slowing  them down and 
requires sophisticated techniques for cooling.
While the standard procedure of trapping and cooling neutral atoms and ions relies on adiabatic processes, 	
 meaning that one first slows or even stops a particle beam before one traps it, we explore here the problem of the trapping of
 a (finite velocity) beam of particles by a sudden switch on of a trap. 
Obviously, this is a highly non-adiabatic and instantaneous process but still quite interesting from an experimental point of view.
Taking particles suddenly out of an atomic beam is an immediate and straightforward procedure, 
which would be a simple alternative to the efficient adiabatic trapping because it
 could be easily implemented in any  experiment with cold atoms. 
Here, we illustrate this sudden trapping mechanism in terms of a classical phase space analysis of the atomic ensemble.

To be specific,
we consider an atomic ensemble of $N=1240000$ atoms, initially in free motion,  
and investigate the trapping features of a crossed optical dipole
trap, which is suddenly turned on. 
Based on the spatial dependence of the optical potential \eqref{poten2}, we
restrict the atomic ensemble spatial extension to the ``prism" ${\cal P}=(-2.5 \le x \le 2.5, -2.5 \le y \le 2.5, -1.5 \le z \le 1.5)$. 
The initial positions of the atoms  are uniformly distributed in this prism  ${\cal P}$.
We assume that this atomic beam is in thermal equilibrium propagating along the $x$ axis ($v_y=v_z=0$),
with a velocity distribution~\cite{MetcalfStraten} 
\begin{equation}
\label{distribution}
f(v_x) = \frac{4 v_x^3}{v_{rms}^4} \exp (-\frac{2 v_x^2}{v_{rms}^2}),
\end{equation}
where $v_{rms}$ is the root mean square beam velocity related to the mean
kinetic energy $K_{rms}= v_{rms}^2/2$, in reduced units.
For a given  mean kinetic energy,  and  using the Monte-Carlo method, an
initial positive velocity $v_x$ is assigned to each atom according to the velocity distribution \eqref{distribution}. 
 
Initially the optical trap is off, and the atoms move freely with  only kinetic energy 
$E= v_{x}^2/2$.
The optical trap is instantaneously switched on at $t=0$, and  the free atomic motion is perturbed by 
the dipole potential.
The initial energy $E= v_{x}^2/2$  decreases by $u(x,y,z)$, which depends on the 
atom position. 
Due to the shape of the potential,  atoms around the origin should be strongly perturbed and likely trapped.

The trajectories of  the  atoms  are computed by integrating
numerically the Hamiltonian equations of motion \eqref{ecumovi}. If after a
convenient propagation cut-off time $t_c$, fixed here to  $t_c=2\times 10^4$,
the distance of an atom from the trap center is still 
below the threshold distance $r_t=5$, we consider that the
atom is  trapped.
% In this paper we take $t_c=2\times 10^4$.
To choose $t_c$, we have taken into account that 
for $E>-0.5$, the chaotic orbits are always escape orbits, see the previous section,
with an elapsed time before leaving the trap being 
extremely long in many cases, \ie, the diffusion is extremely slow compared to the trapping time.
For instance, in a typical dipole trap for Rb atoms with beam waist $\omega_o=25 \ \mu$m and
well depth $U_o = 1.5$~mK~\cite{wieman}, 
the frequency  and time units are
$\nu=(U_o/w_o^2 m)^{1/2} \approx 1.5 \times 10^{4}$~s$^{-1}$ and  
$t\approx 1/1.5 \times 10^{-4}$~s, respectively.
Since in a conventional experiment, the trap is filled in a few seconds, the cut-of time value
$t_c=2\times 10^4$ seems to be appropriate. Note that we are assuming as trapped orbits
the chaotic ones whose escape time is above $t_c=2\times 10^4$.

\begin{figure*}[t]
\centerline{
\includegraphics[scale=0.28]{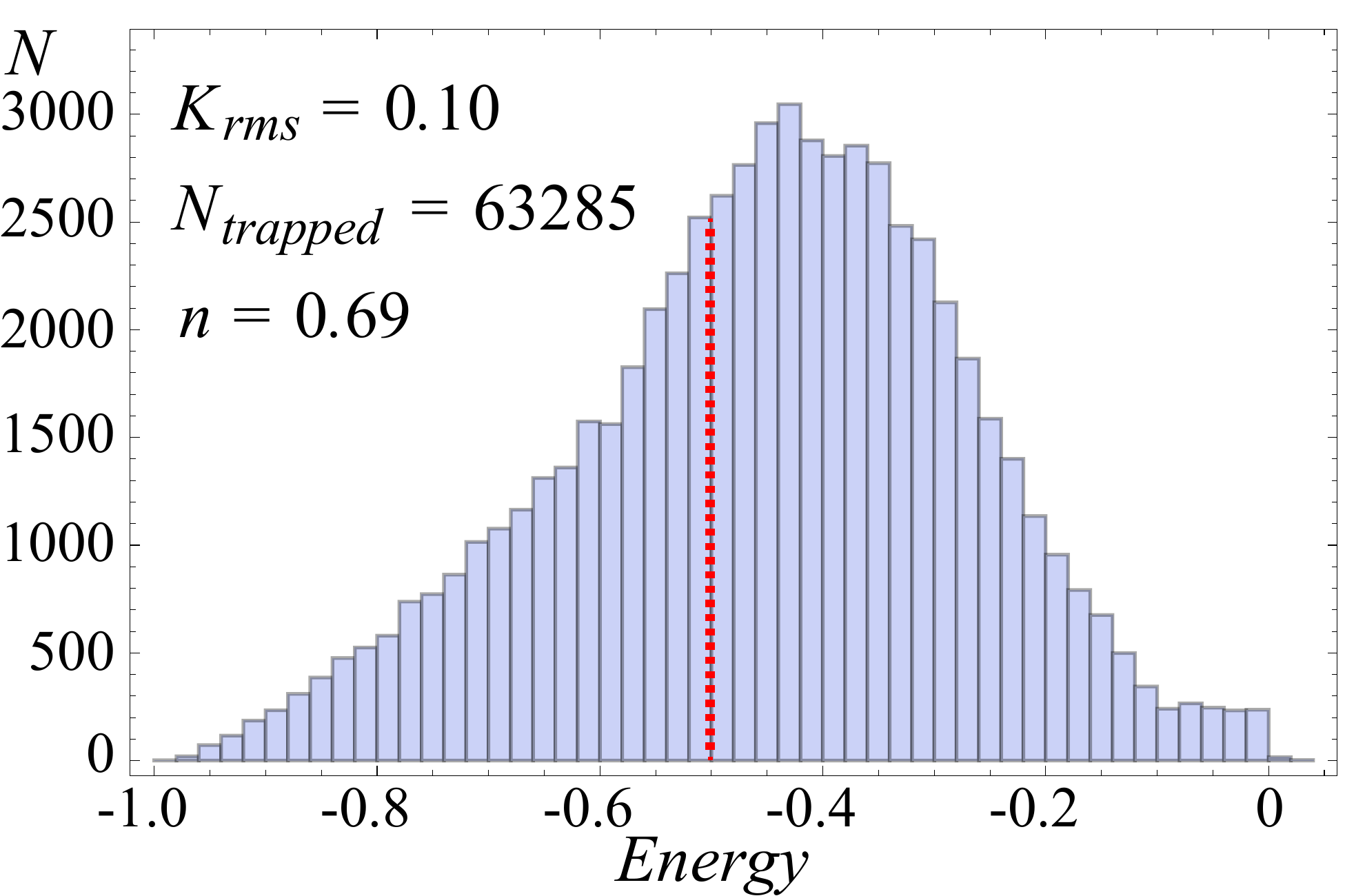}
\includegraphics[scale=0.28]{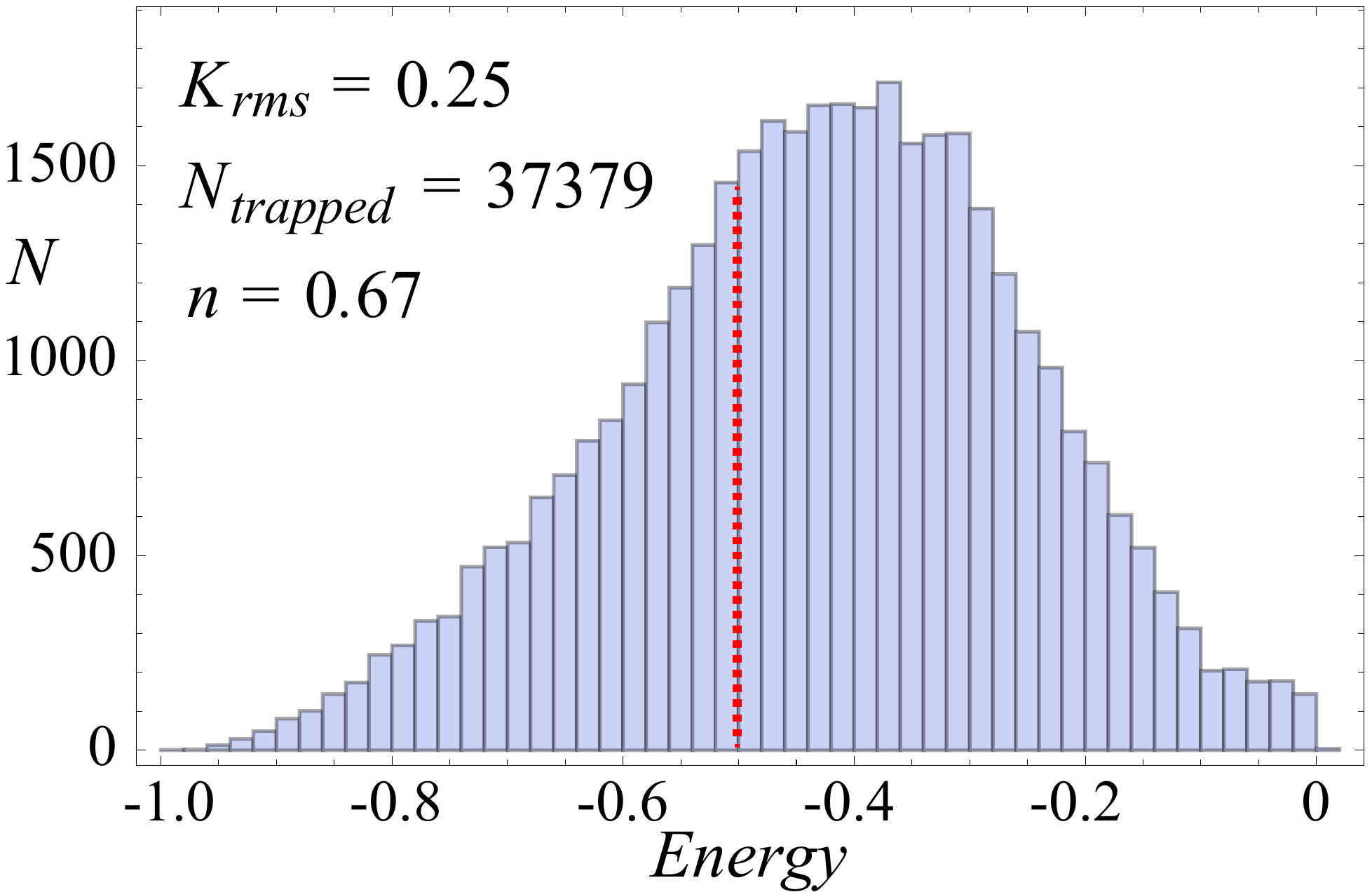}
  \includegraphics[scale=0.275]{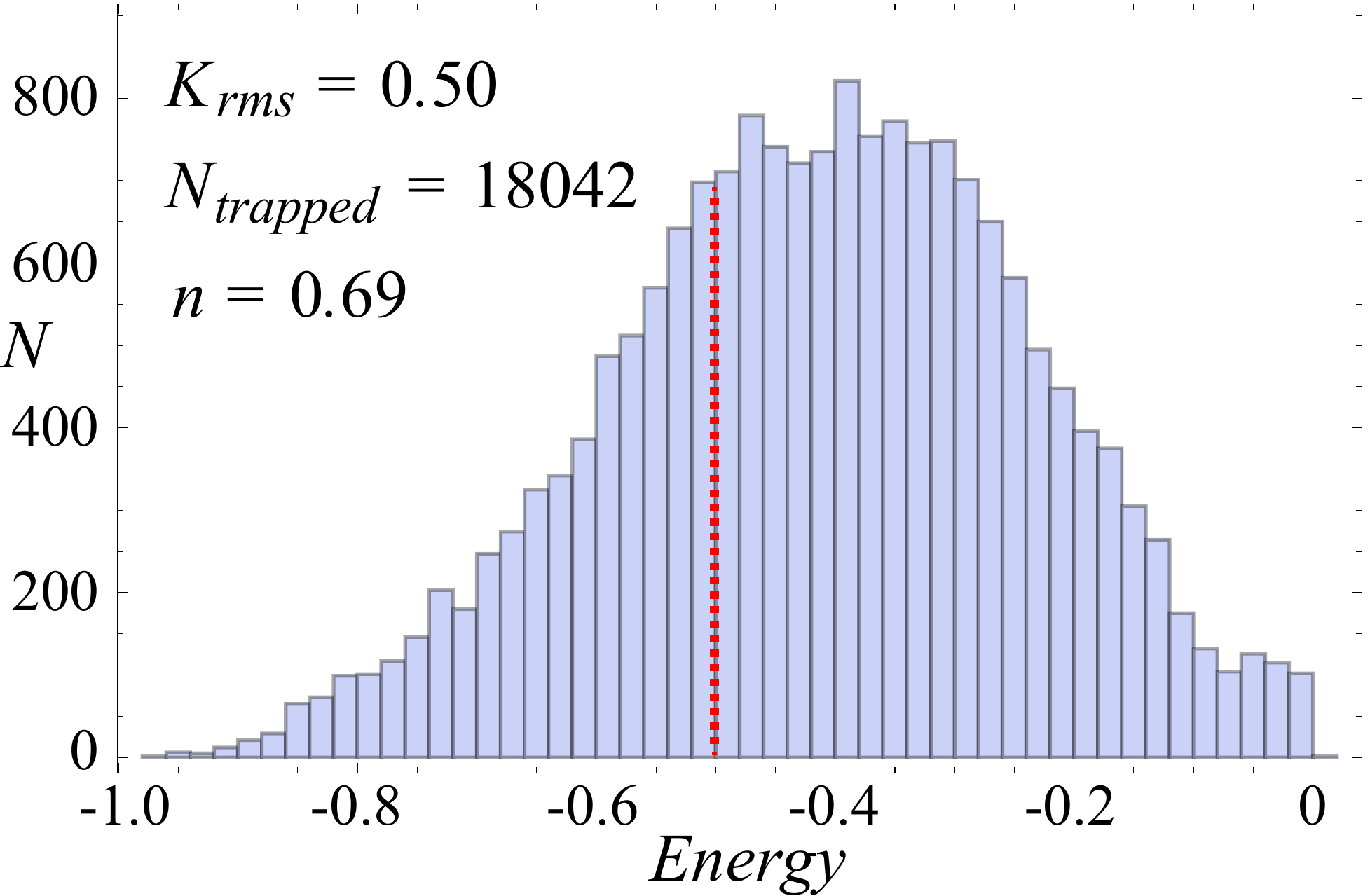}
} 
\bigskip
\centerline{
\includegraphics[scale=0.28]{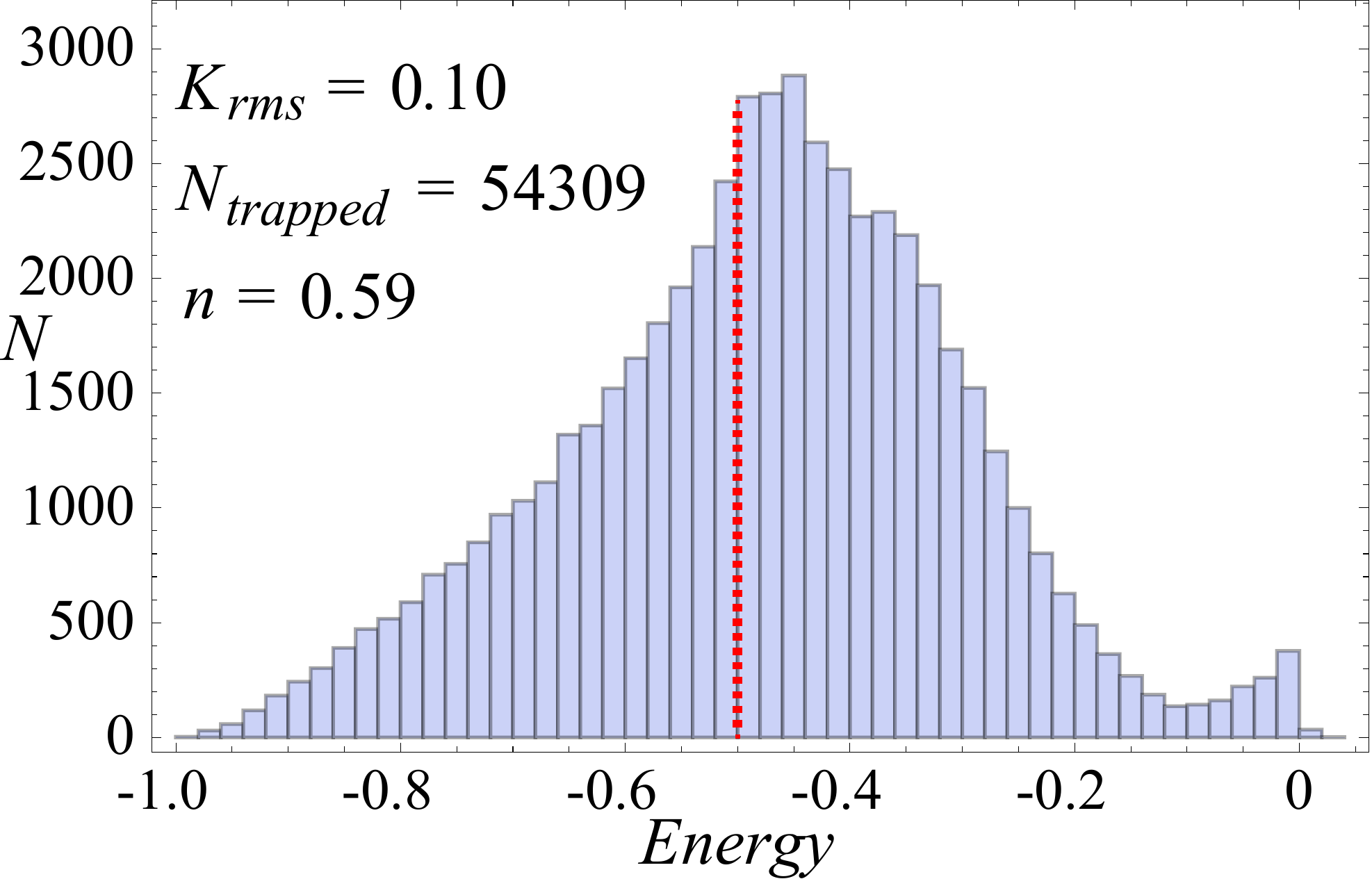}
\includegraphics[scale=0.28]{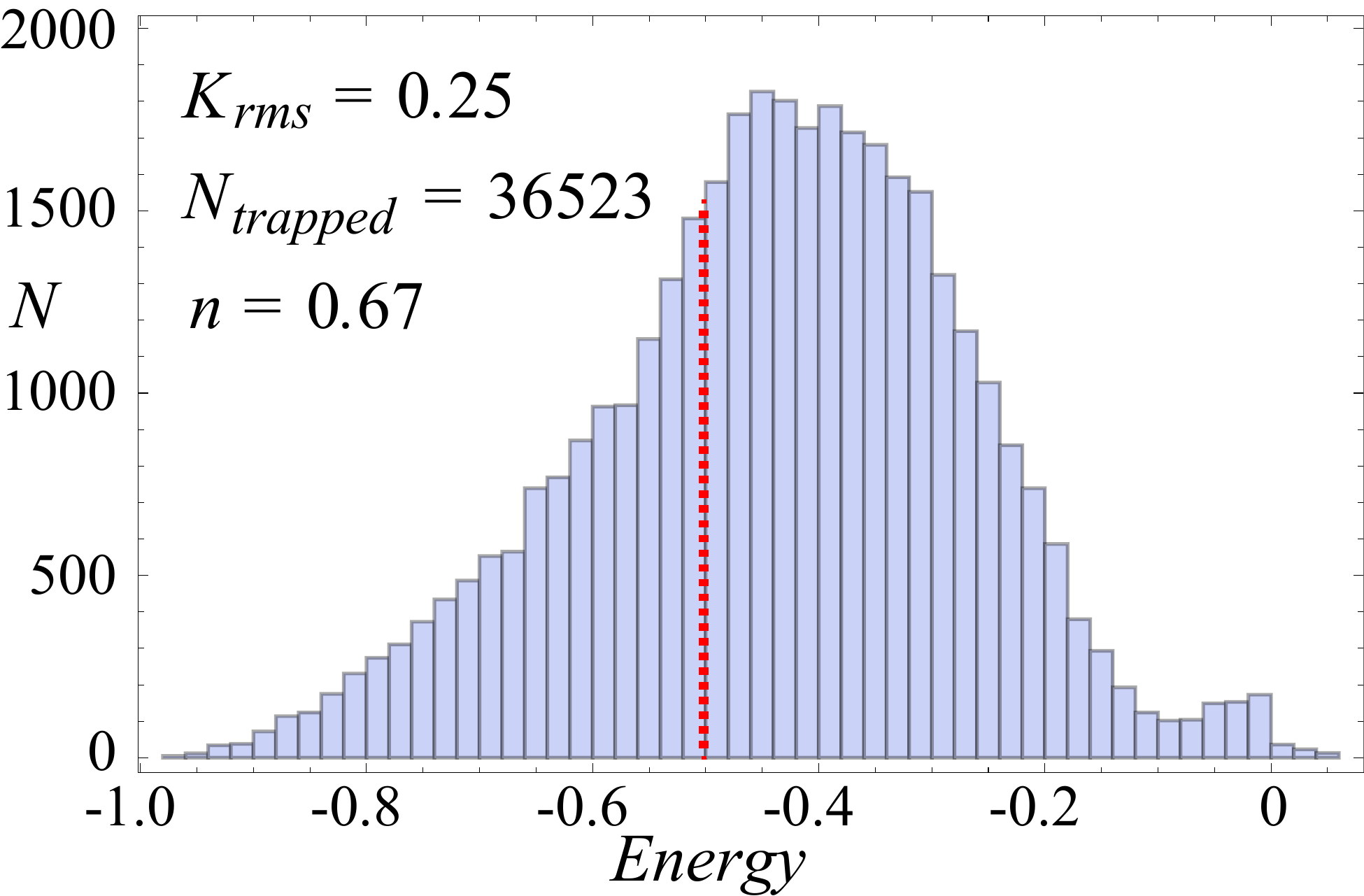}
\includegraphics[scale=0.28]{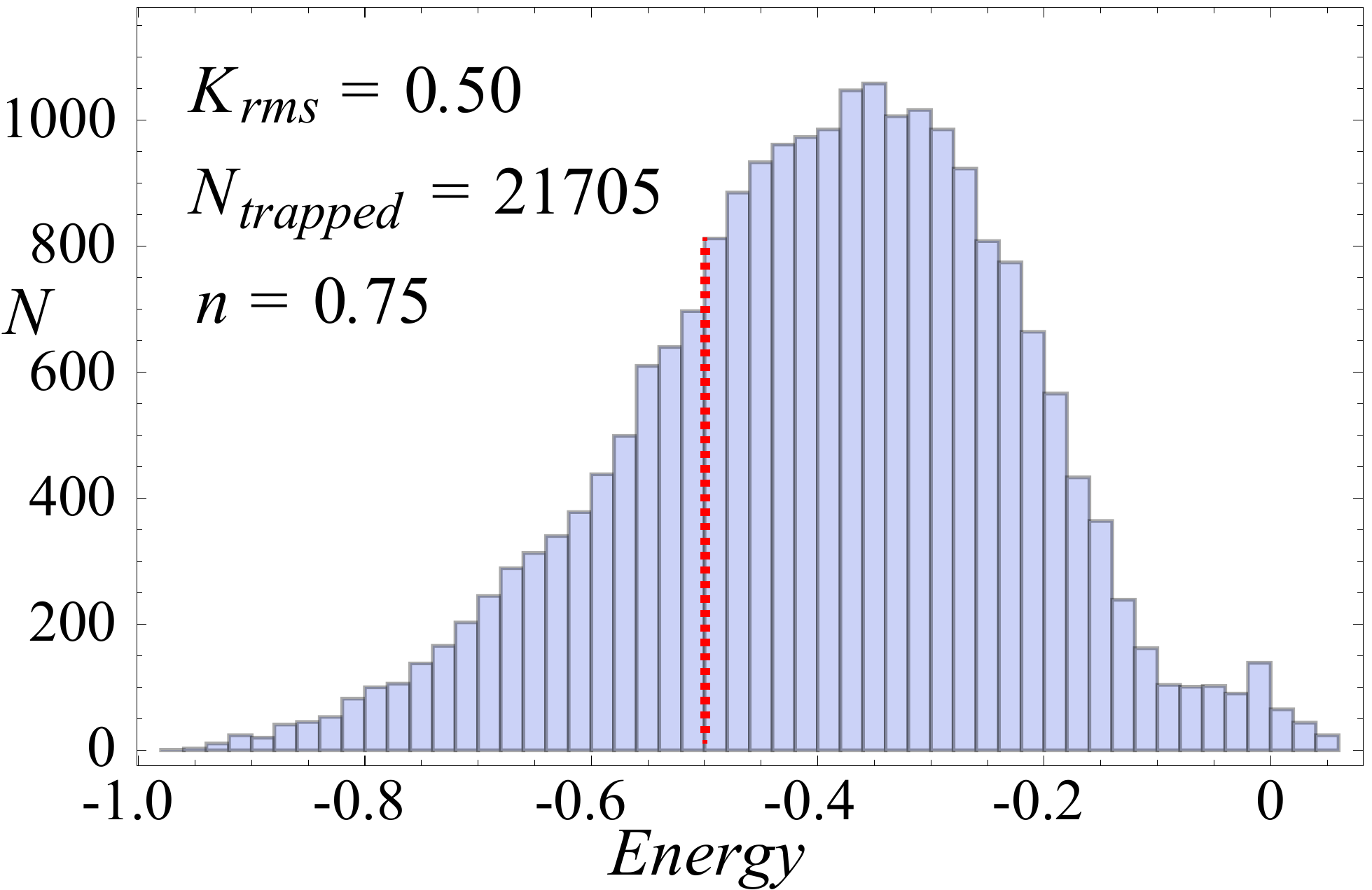}}
\caption{Energy distribution histograms of the trapped orbits for three different 
mean kinetic energies $K_{rms}$ of the atomic beam.  $N_{trapped}$ is the
number of trapped atoms and $n$ is  the ratio  of the number of trapped atoms
with energy greater than $E=-0.5$ and  $N_{trapped}$. Upper and lower row histograms correspond to
the case of the atomic beam propagating along the $x$ and $z$ axes, respectively.}
\label{fi:trajectories3D}
\end{figure*}

For three mean kinetic energies $K_{rms}$ of the atomic beam, 
we present in \autoref{fi:trajectories3D} the energy distribution histograms of the trapped orbits,
indicating the number of trapped atoms $N_{trapped}$ 
and the ratio $n$ between the number of trapped atoms
with energy larger than the threshold energy
($E>-0.5$) and $N_{trapped}$.
As it should be expected, by increasing the 
initial  mean kinetic energy $K_{rms}$ the amount of trapped atoms $N_{trapped}$ decreases.
For the fast beam with $K_{rms}=0.5$, only $1.4\%$ of the initial atoms remain trapped. 
However, the majority of these atoms have an energy larger than the escape channels, indeed,  $n$  is always larger than $0.65$, \ie, for more than $65\%$ of the trapped atoms, $E>-0.5$. This is also observed in the large tail of the histograms in
Fig.\ref{fi:trajectories3D}.

Now, we compare the trapping ability of the optical trap 
for atomic beams propagating along the $x$ and $z$ directions.
We consider  that the atoms move in the $z$ axis  with velocity  $v_z$ given by the
 distribution \eqref{distribution} and $v_x=v_y=0$.
The corresponding results are presented in the  histograms of the lower panels in  \autoref{fi:trajectories3D}, where we observe 
 quantitatively similar results. As $K_{rms}$ is increased, $N_{trapped}$ decreases whereas $n$ increases. 
 In particular,  for $K_{rms}=0.10$ and $0.5$, the percentage of trapped atoms with $E>-0.5$ are $59\%$ and $75\%$, respectively.
Hence, we can conclude that  the trapping ability of the optical trap is
very similar for an atomic beam propagating along the $z$ and $x$ axes.

The explanation of why such a large  amount of
atoms with $E>-0.5$ remains trapped can be found in the OFLI$^{TT}_2$ maps. 
In the $E=-0.3$   OFLI$^{TT}_2$ maps of  \autoref{fi:ofli2_030201}~(a) and (b), 
the black points indicate the initial conditions of the atomic beam propagating along the $x$
and $z$ axes,
 respectively, which are confined when the optical trap is switched on. 
In both cases,  most of the black
points lay on green  regions of bounded
regular motion. Thus, after switching on the lasers,  the confined atoms with
energy $E>-0.5$ are in phase space regions where bounded motion is still possible due
to the existence of persistent KAM tori. Indeed, these dynamical structures are responsible
for the capture of atoms with an energy larger than the trapping energy threshold. 
There are also many black points in  red chaotic regions, 
which are  escape chaotic orbits that remain trapped for long periods of time.
For $E=-0.2$, the  initial conditions of the confined orbits are also plotted 
 in    \autoref{fi:ofli2_030201}~(c) and (d). 
In this case, most of the bounded trajectories are confined in chaotic regions because
the regular KAM tori regions have shrunken.

\section{Conclusions}
\label{sec:conclusions}

We have investigated the nonlinear dynamics of an atom in a crossed optical dipole
trap formed by two identical Gaussian laser beams propagating along perpendicular directions. 
 The evolution of the stability of the dynamics  with increasing energy has been shown by 
 a detailed analysis of the 
phase space in terms of two dimensional OFLI$^{TT}_2$ maps. 
For small energies, the phase space is populated with periodic and quasi periodic orbits, but  
 chaotic motion appears as the energy is increased.
 Above a certain threshold, escape in both spatial ($x$, $y$) degrees of freedom becomes possible and   
the dynamics is of mixed regular and chaotic character. 
Regular trajectories in which  the atom quickly leaves the trap, and 
chaotic ones with very long escape times exist in this case. 
For energies close to zero,  the threshold for escape in all three spatial directions, 
the phase space still presents small areas of trapped regular motion which are surrounded by a 
 sea of  chaotic scattering~\cite{ott}.

Furthermore, we have explored the impact of an optical dipole 
trap, which is suddenly turned on, on an atomic beam that moves freely. 
Independently of the initial direction of the atoms, we encounter that some of them remain within the trap, 
and the amount of trapped atoms decreases, as expected,  as the initial atomic energy is increased.   
 The majority of these trapped atoms have an energy larger than the escape 
channels, which can be understood in terms of the phase space structures. 
 The  OFLI$^{TT}_2$ maps show that these trapped atoms are either  in bounded orbits, which are possible due
to the existence of KAM tori, or in chaotic ones with very long escape times.  

A natural extension of this work would be to investigate the dynamics following a  quench. 
For an atomic ensemble in an optical dipole trap,  a sudden
change of the trap depth or  width would  provoke significant changes in the phase space structure. 
A classical study based  on OFLI$^{TT}_2$ maps would then characterize the 
escape dynamics and trapped population.

\begin{acknowledgments}

R.G.F. gratefully acknowledges a Mildred Dresselhaus award from the excellence cluster 
"The Hamburg Center for Ultrafast Imaging Structure, Dynamics and Control of Matter
at the Atomic Scale" of the Deutsche Forschungsgemeinschaft. R.G.F. 
also acknowledges financial support by the Spanish project FIS2011-24540 (MICINN), the
Grants P11-FQM-7276  and FQM-4643 (Junta de Andaluc\'{\i}a), and Andalusian 
research group FQM-207. M.I. and J.P.S. acknowledge financial support by the Spanish project
MTM2011-28227-C02-02 (Spanish Ministry of Education 
and Science).

\end{acknowledgments}


\begin{thebibliography}{}
\bibitem{footnote_mia}  Strictly speaking,   
an orbit is chaotic if its largest Lyapunov exponent is positive, which implies a sensitive dependence on initial conditions.
A chaotic orbit in a close system is characterized by the finiteness of the Liapunov exponent, as a consequence of which 
the exponential sensitivity has its trace at all phases of the chaotic orbits (we exclude here, for simplicity, the case of stickiness to
regular islands).
In contrast, for the chaotic scattering in an open system, there is an asymptotic ballistic motion, that 
is distinguished from the chaotic behavior in a closed system.
\end{thebibliography}
\end{document}